\documentclass[useAMS, usenatbib]{mn2e} 

\usepackage{graphicx}

\title{Time-dependent particle acceleration in supernova remnants in different environments}
\author[K.M. Schure et al.]{K.M. Schure$^{1}$\thanks{E-mail: K.M.Schure@phys.uu.nl}, A. Achterberg$^{1}$, R. Keppens$^{1,2,3}$, and J. Vink$^{1}$\\
$^{1}$Astronomical Institute, Utrecht University,Postbus 80000, NL-3508 TA Utrecht, The Netherlands\\
$^{2}$Centre for Plasma Astrophysics, K.U. Leuven, Belgium\\
$^{3}$FOM-Institute for Plasma Physics ``Rijnhuizen'', Nieuwegein, The Netherlands}

\begin{document}

\newcommand\araa{{ARA\&A}}
\newcommand\apj{{ApJ}}
\newcommand{\apjl}{ApJL}
\newcommand\apjs{{ApJS}}
\newcommand\aap{{A\&A}}
\newcommand\mnras{{MNRAS}}
\newcommand\rmxaa{{Rev. Mexicana Astron. Astrofis.}}
\newcommand\nat{{Nature}}
\newcommand\physrep{{Phys.~Rep.}}

\date{\ldots; \ldots}
\pagerange{\pageref{firstpage}--\pageref{lastpage}}\pubyear{2010}
\maketitle
\label{firstpage}

\begin{abstract}
We simulate time-dependent particle acceleration in the blast wave of a young supernova remnant (SNR), using a Monte Carlo approach for the diffusion and acceleration of the particles, 
coupled to an MHD code. 
We calculate the distribution function of the cosmic rays concurrently with the hydrodynamic evolution of the SNR, and compare the results with those obtained using simple steady-state models.
The surrounding medium into which the supernova remnant evolves turns out to be of great influence on the maximum energy to which particles are accelerated. 
In particular, a shock going through a $\rho \propto r^{-2}$ density profile causes acceleration to typically much higher energies than a shock going through a 
medium with a homogeneous density profile. We find systematic differences between steady-state analytical models and our time-dependent calculation in terms of spectral slope, 
maximum energy, and the shape of the cut-off of the particle spectrum at the highest energies. 
We also find that, provided that the magnetic field at the reverse shock is sufficiently strong to confine particles, cosmic rays can be easily re-accelerated at the reverse shock.
\end{abstract}


\section{Introduction}
\label{sec:intro}
Cosmic rays with energies up to at least $\sim 10^{15}$~eV are thought to originate in supernova remnants (SNRs). They are high-energy particles that have a simple power-law energy spectrum that extends over five decades in energy. The need for an efficient acceleration mechanism in SNRs has motivated the development of the theory of diffusive shock acceleration (DSA), according to which particles are accelerated at shock fronts \citep[see~][for a comprehensive review]{2001MalkovDrury}. 

Young supernova remnants are ideal locations for studying this process, because of the high shock velocity, and the presence of a few nearby young remnants for which detailed observations are available \citep[e.g.~][]{2004Hwangetal, 2005Bambaetal, 2009Aceroetal}. The presence of high-energy electrons spiralling in a $\sim 10\ \umu$G magnetic field has been established from the emission of synchrotron radiation from radio wavelengths to X-rays. Both the magnetic field strength and the typical particle energy are inferred from synchrotron theory and can be used to compare with theoretical predictions \citep{1994Achterbergetal,2003VinkLaming, 2005Voelketal, 2005Vink}. Even though synchrotron emission only indicates the presence of relativistic electrons, the presence of energetic protons is suggested by the observations of TeV gamma rays \citep[e.g.~][]{2009Aharonianetal}, and by indications of magnetic field amplification beyond that what is expected from simple shock compression \citep{2005Warrenetal, 2008CassamChenaietal}. In addition, there are indications that the SNR blast waves are not simple hydrodynamic blast waves in a single-component gas. They behave in a way that indicates that a significant fraction of the pre-shock energy density resides in cosmic rays \citep{2000Decourchelleetal,2009Helderetal}. 

Various groups work on trying to get an integral picture of the interaction between SNR shocks, magnetic fields, and the associated particle acceleration \citep[e.g.~][]{1999BerezhkoEllison,2005KangJones,2005AmatoBlasi, 2006Vladimirovetal,2006BerezhkoVoelk,2007ZirakashviliAharonian,2010Ferrandetal}. The difficulty is the fact that the process is inherently non-linear. 
The spectral slope of the particles is determined mainly by the difference between the plasma velocities on the two sides of the shock. This difference depends on the compression ratio, which in turn is determined by the effective equation of state of the gas-cosmic ray mixture: the presence of cosmic rays tends to soften the equation of state around the shock, which in turn increases the total compression. 
In addition, the gradient in cosmic ray pressure slows down the incoming flow in the cosmic-ray precursor to the shock. High-energy particles with a larger scattering mean free path probe a larger region around the shock and `feel' a higher compression ratio. For these reasons the spectrum flattens at the high-energy end in fully non-linear models. An additional nonlinearity arises when the magnetic fields are amplified by the streaming of cosmic rays.  

Cosmic rays isotropise by scattering off Alfv\'en waves. In gyro-resonant scattering the scattering rate depends on the cosmic ray energy through the slope of the spectrum of the scattering waves. It is often assumed that the diffusion rate is described by Bohm diffusion, where the diffusion coefficient scales as $\kappa_{\rm B} \propto E$, with $E$ the energy of the particle. These waves are self-generated by the streaming of cosmic rays away from the shock, through a resonant instability \citep[e.~g.][]{1975Skilling}, and / or through a non-resonant instability \citep{2001BellLucek, 2004Bell, 2009LuoMelrose}.

Various authors focus on the feedback of cosmic rays onto the hydrodynamics near the shock \citep{1997Malkov, 2007Blasietal, 2009Patnaudeetal,2010Ferrandetal}. The distribution function of the particles is calculated, from which the cosmic ray pressure can be determined. The pressure term alters the equation of state and feeds back on the hydrodynamics. Alternatively, a standard power law distribution is assumed, simplifying and speeding up the process, making it fast enough for application on larger scales \citep{2007Ensslinetal}. The disadvantage of this approach is that -- generally speaking-- the time-dependence or the influence of a complicated magnetic field geometry can not be taken into account.

Our work focuses on the time- and position dependence of cosmic ray spectra in the supernova remnant, where we use the test-particle approximation, neglecting feedback of the particles onto the plasma. The high-energy particles isotropise on both sides of the shock, due to scattering off waves, and are accelerated by repeated crossing cycles at the shock. The acceleration of relativistic particles can be described by a set of stochastic differential equations (SDEs) \citep{1992AchterbergKruells,1994KruellsAchterberg} and this has been applied succesfully by a number of authors  \citep{1999MarcowithKirk, 2004vanderSwaluwAchterberg, 2010MarcowithCasse} in various hydrodynamic codes. 

We use the adaptive mesh refined magneto-hydrodynamics (MHD) code: AMRVAC \citep{2003Keppensetal, 2007HolstKeppens} as the framework for our particle acceleration method.
Not only do we calculate the acceleration/deceleration of test particles due to compression/expansion of the flow, we also model the dependence of diffusion and radiative losses, while keeping computational costs down with the adaptive mesh strategy. Our approach has the advantage that it is able to also tackle a more complicated circumstellar density profile 
than other models. The disadvantage is (for now) having to neglect the nonlinear feedback of the cosmic rays onto the plasma. 

We describe the different models we use to calculate the particle spectrum in supernova remnants in \S~\ref{sec:1Dmodels}. In \S~\ref{sec:theory} we will discuss the theory of diffusive shock acceleration and the evolution of the supernova remnant, and derive some analytical estimates for the expected cosmic ray spectrum. The method and set-up of the simulations will be described in \S~\ref{sec:method}. In \S~\ref{sec:testmodels} we will describe some test models and results obtained with this method. We will subsequently present the results for the particle spectra from the SNR models in \S~\ref{sec:snr} and conclude with a discussion and summary in \S~\ref{sec:discussion}.

\section{SNR models}
\label{sec:1Dmodels}

We simulate the evolution of the SNR and concurrently calculate the cosmic ray distribution in phase-space for the various models. In these models we evaluate the effect of: {\em i}): the approximation that is used for the diffusion coefficient, {\em ii}): the adopted equation of state, 
{\em iii}): the density profile of the background into which the supernova remnant evolves, and {\em iv}): the strength of the magnetic field, both in the surrounding medium, and in the ejecta.  
Further details of the latter three can be found below.
We summarize the entire set of models used in our simulations in Table~\ref{table:models}.

\subsection{Influence of the equation of state}

The compression ratio at the shock depends on the adiabatic index (specific heat ratio) of the plasma. 
Since the expected slope of the spectral energy distribution of the cosmic rays depends on the compression ratio, by varying the equation of state we can test our code and see if the results change according to expectation.

Besides providing a nice test for the code, varying the adiabatic index is also physically relevant for systems in which efficient cosmic ray acceleration occurs. The adiabatic index of the plasma parametrises the equation of state of the plasma. The value of $\gamma = 5/3$ corresponds to the case where the gas consists of a mono-atomic gas, and is what we use for simulations in which the contribution of cosmic rays is neglected. The value of  $\gamma = 4/3$ corresponds to the adiabatic index of a relativistic gas. A plasma in which a large part of the pressure is made up by cosmic rays has an effectively softer equation of state, with a value of $4/3 <  \gamma < 5/3$. When cosmic rays escape from the system, they may drain the shock region of energy, softening the equation of state further.

Approximating the hydrodynamics with a lower, but fixed, adiabatic index is only a first step to see how the slope of the cosmic ray spectrum is affected by a softer equation of state. 
In reality, the cosmic ray pressure gradient in the shock precursor slows down the incoming flow, which in turn results in a concave cosmic ray spectrum if the scattering mean free path increases with energy. Additionally magnetic field amplification may partly reverse the effect by stiffening the equation of state. 

\subsection{The explosion environment}

We consider two situations for the environment into which we let the ejecta expand. 
{\em i}): The case of a homogeneous medium, such as may be applicable to Type Ia SNe. These models will be referred to as interstellar medium (ISM) models. 
{\em ii}): A $\rho \propto R^{-2}$ profile for the plasma, as will arise from a steady stellar wind. This situation is more applicable to core collapse supernovae whose
progenitors have strong stellar winds. These models will be referred to as circumstellar medium (CSM) models.
We set the density of the homogeneous ISM to a value of $2.34 \times 10^{-24}$~g~cm$^{-3}$. 
The wind parameters for the $\rho \propto R^{-2}$ CSM are chosen such that the volume containing an amount of mass equal to the ejecta mass is equal in both cases. 
The ejecta mass that we use is $3.5$~M$_\odot$ (Sect.~\ref{sec:ejecta}). The radius containing an equal amount of mass in the ISM is $R_{3.5 {\rm M}_\odot}=\left(3 M/4 \pi \rho\right)^{1/3} \sim 3$~pc. 
We employ mass loss parameters that are typical for the winds from red supergiants,  with a value for the mass loss rate of $\dot M=1.71\times10^{-5}$~M$_\odot$~yr$^{-1}$ 
and a wind velocity equal to $v_{\rm wind}(10^{16} {\rm cm})=10$~km~s$^{-1}$. In all cases we set the initial temperature of the CSM/ISM and the ejecta to $T=10^4$~K.

\subsection{Magnetic field}

For the magnetic field we look at the case of a constant magnetic field strength, with values of $3$, $10$, and $20$ $\umu$G. 
We restrict ourselves for now to parallel shocks, where the magnetic field is aligned with the shock normal, for reasons mentioned in \S~\ref{sec:method}. 
In this case the magnetic field is the same on both sides of the shock and the diffusion coefficient depends only on the particle energy in the case of Bohm diffusion. 
While the code has an MHD solver and we can follow the magnetic field dynamically, this only becomes interesting in two dimensions, something we defer until a later work. 
We solve the Euler equations and parametrise the magnetic field strength for use in the SDEs.

We find (see \S~\ref{sec:rev}) that particles diffuse far enough downstream to reach the reverse shock such that they are re-accelerated, provided that the magnetic field in the ejecta is strong enough to confine the particles. 
We therefore investigate two different cases for the magnetic field in the ejecta: we either parametrise the magnetic field strength such that it decays with radius as $R^{-2}$ as to represent the expanding fossil field of the progenitor, frozen into the plasma while conserving magnetic flux, 
or we keep the magnetic field fixed at the same value as in the ISM/CSM. In the first case, the magnetic field strength for reasonable progenitor fields is very weak, making the gyroradius (and thus the diffusion length-scale) of the particles very large when Bohm diffusion is assumed. 
In the latter case, the diffusion in the ejecta proceeds at a similar rate as in the ISM/CSM, so particles that encounter the reverse shock can be re-accelerated, as we will illustrate in Sec.~\ref{sec:rev}.

\begin{table*}
\centering                 
\caption{\textrm{Overview of the supernova remnant models.} }           
\label{table:models}        
\begin{tabular}{c  c  c   c   c  c  c}       
\hline\hline
MODEL: & ISM$\kappa_{\rm c}$ & ISM$\kappa_{\rm B}$ & CSM$\kappa_{\rm B}$ & ISM$\kappa_{\rm B}$r & CSM$\kappa_{\rm B}$r & CSM$\kappa_{\rm B}$s
\\
\hline
$\rho$ background (g~cm$^{-3}$)& $2.34\times 10^{-24}$ & $2.34\times 10^{-24}$ & $\propto r^{-2}$& $2.34\times 10^{-24}$ & $\propto r^{-2}$ & $\propto r^{-2}$\\
diffusion coefficient & constant & Bohm & Bohm & Bohm & Bohm & Bohm\\
magnetic field strength ($\umu$G) & $10$ &  3, 10, 20 & 3, 10, 20 & 20 & 20 & 20 \\
reverse shock acceleration & no & no & no & yes & yes & no\\
equation of state & normal ($\gamma=5/3$) & normal & normal & normal & normal & soft ($\gamma=4/3$)\\
\hline                                   
\end{tabular}

\end{table*}

\section{Theory}
\label{sec:theory}
\subsection{Diffusive Shock Acceleration}
At the forward shock of SNRs, and possibly also at the reverse shock \citep{2008HelderVink, 2010ZirakashviliAharonian}, 
particles are believed to be accelerated to relativistic energies,  up to $10^{15}$~eV for protons. 
The favoured mechanism is diffusive shock acceleration (DSA), where every time the particle crosses the shock interface it experiences a Lorentz boost, after which the momentum direction is randomised in the local frame. Since the plasma rest frames on either side of the shock differ in speed, this results in a systematic increase of the particle momentum every time the particle cycles from upstream to downstream and back \citep{1978Bella}. 

\citet{1977Krymskii, 1977Axfordetal, 1978Bella,1978Bellb, 1978BlandfordOstriker, 1983Drury} derive the spectrum of shock-accelerated particles, based on the energy gain per cycle versus the probability of being advected away and lost from the acceleration process. The escape probability is 
given by $P_{\rm esc}=4 V_{\rm s}/rc$, where $V_{\rm s}$ is the shock velocity, $r$ the shock compression ratio and $c$ is the velocity of the cosmic rays, equal to the speed of light. 
The mean time $\Delta t$ for a relativistic particle to complete a cycle probing the upstream and the downstream plasma and the corresponding boost in momentum $p$ is given by \citep{1983LagageCesarsky,1983Drury}:

\begin{eqnarray}
\Delta t =\frac{4}{c V_{\rm s}}(\kappa_1+r \kappa_2),\\
\Delta p =\left(\frac{r-1}{3r}\right)\frac{4V_s}{c}p.
\end{eqnarray}
where $\kappa$ is the diffusion coefficient and we will use the subscript $1$ for upstream-, and $2$ for downstream properties. 

The test particle limit results in a power-law spectrum in momentum, with the number density of particles per unit momentum following the distribution
\begin{eqnarray}
F(p)\equiv\frac{\partial {\cal N}}{\partial p}\propto p^{-q},
\end{eqnarray}
with
\begin{eqnarray}
\label{eq:q}
q=1+\frac{\ln(1/P_{\rm ret})}{\ln((p+\Delta p)/p)}=\frac{r+2}{r-1}
\end{eqnarray}
the slope. Here $P_{\rm ret}=1-P_{\rm esc}$ is the return probability in a shock crossing cycle.

\subsection{Acceleration time-scales}
\label{sec:time-scalesanalytical}

The acceleration rate of relativistic protons or electrons of given energy $E = pc \gg m_{\rm p}c^2\sim 1$~GeV is:
\begin{eqnarray}
\label{eq:accrate}
\left(\frac{{\rm d}E}{{\rm d}t}\right)_{\rm dsa}=\left(\frac{r-1}{3r}\right)\frac{V_{\rm s}^2}{(\kappa_1+r\kappa_2)}E,
\end{eqnarray}
giving a typical time-scale for acceleration time of
\begin{eqnarray}
t_{\rm acc} \equiv \left(\frac{1}{E} \frac{dE}{dt}\right)^{-1} = 
\frac{3r(\kappa_1+r\kappa_2)}{(r-1)V_{\rm s}^2}\label{eq:tacc}.
\end{eqnarray}

It will be useful to consider the case of Bohm diffusion for comparison with the simulations, and to look at the combined effect of acceleration and synchrotron losses, which is important for electrons. The change in energy at the shock is then given by: 
\begin{eqnarray}
\label{eq:lossrate}
\left(\frac{{\rm d}E}{{\rm d}t}\right)_{\rm acc+synch}=\frac{3 Z e B V_{\rm s}^2}{(1+r)r c}-\frac {\lambda_s B^2 E^2}{m c^2},
\end{eqnarray}
for a parallel shock, where $\kappa_1=\kappa_2=D_{\rm B} = c E/3ZeB$ is the Bohm diffusion coefficient, 
which depends on the gyroradius of the particles, $B$ the magnetic field strength, $Z$ the charge number, $e$ the elementary electric charge and $m$ the mass of the particle.

We have introduced a term that accounts for synchrotron losses with
\begin{eqnarray}
\label{eq:betas}
\lambda_s=\frac{\sigma_T}{6 \pi m c} \propto m^{-3}.
\end{eqnarray}
The mass dependence makes this term negligible for protons. High-energy electrons on the other hand lose a substantial fraction of their energy by synchrotron radiation. 
With a Thomson cross section of $\sigma_T=6.65\times10^{-25}$~cm$^2$ for electrons this gives for $\lambda_s=1.29\times10^{-9}$~cm~s~g$^{-1}$. 
The same loss-term can be used to account for inverse Compton losses, where the energy that is lost in upscattering photons from the microwave background corresponds 
to synchrotron losses in an equivalent magnetic field $B_{\rm CMB}=3.27\ \umu$G \citep{1998Reynolds}. When the actual magnetic field is stronger than this value, synchrotron losses will dominate over inverse Compton losses unless there is another source of photons that boosts the inverse Compton scattering rate. For now we neglect Compton losses, which means that we slightly over-estimate the maximum energy
for electrons in our models with $B \sim 3 \: \mu{\rm G}$.

For oblique shocks the compression of the magnetic field and the change in residence times upstream and downstream have to be taken into account.

\subsection{Maximum energy of the cosmic rays}

The maximum attainable energy $E_{\rm max}$ for cosmic rays depends on the size of the accelerator, 
the time they have spent there, and the strength of adiabatic- and synchrotron losses. In order to determine this, we follow the evolution of the supernova remnant. Initially, the ejecta expand freely into the surrounding medium. The expansion slowly decelerates as the ejecta sweep up mass from the CSM. The deceleration of the blast wave drives a reverse shock into the ejecta, heating the material to millions of Kelvin and making the SNR prominent in X-rays. 
By the time the swept-up mass equals the ejecta mass the Sedov-Taylor phase of SNR evolution begins. 

\begin{figure}
  \centering
 \includegraphics[width=0.5\textwidth]{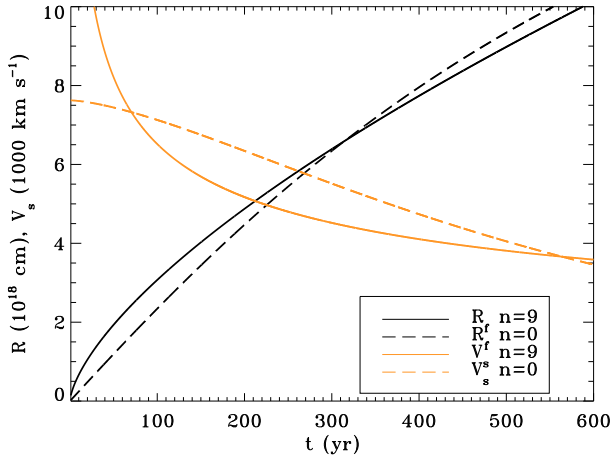}

 \caption[ ] {Early-time evolution of the radius (black) and velocity (coloured) of a SNR in a homogeneous ISM. 
The solid line corresponds to the $n=9$ model from \citet{1999TrueloveMcKee} and the dashed line to the $n=0$ model. 
    \label{fig:vxshock}}
  \end{figure}

Energy conservation gives us typical length- and time-scales, 
and determines the deceleration radius where the transition from free expansion phase to the Sedov-Taylor phase takes place \citep{1959Sedov,1995McKeeTruelove}.
Figure~\ref{fig:vxshock} shows the analytical solution for the evolution of the blast wave of a supernova remnant derived by \citet{1999TrueloveMcKee} . 
We show two different cases, one in which the density of the ejecta is constant with radius ($n=0$-model), and the case in which the ejecta consists of a constant density core with a envelope in which the density decreases with radius as $R^{-9}$ ($n=9$-model, see also our description of the ejecta in the hydrodynamics in \S~\ref{sec:method}). 
The early evolution of SNR radius and blast wave velocity is quite different in the two cases.
The evolution of the blast wave radius and velocity according to \citet{1999TrueloveMcKee} for a $n=9$ power law ejecta envelope into a $1$~cm$^{-3}$ ISM is given by:
\begin{eqnarray}
\label{eq:truelovemckee}
\tilde R_{\rm s}(\tilde t<\tilde t_{\rm ST})&=&1.12\ \tilde t^{2/3}\\\nonumber
\tilde R_{\rm s}(\tilde t>\tilde t_{\rm ST})&=&1.42\ (\tilde t-0.297)^{2/5}\\\nonumber
\tilde v_{\rm s}(\tilde t<\tilde t_{\rm ST})&=&0.75\ \tilde t^{-1/3}\label{eq:shockanalytical}\\\nonumber
\tilde v_{\rm s}(\tilde t>\tilde t_{\rm ST})&=&0.569\ (1.42\ (\tilde t-0.297)^{-3/5}),
\end{eqnarray}
with $\tilde R=R/R_{\rm ch}$ and $\tilde v=v/v_{\rm ch}$. The characteristic radius and velocity are given by 
$R_{\rm ch}=3.07 (M_{\rm ej}/{\rm M}_\odot)^{1/3}$~pc, with $M_{\rm ej}$ the ejecta mass, and $v_{\rm ch}=R_{\rm ch}/t_{\rm ch}$ with $t_{\rm ch}=423 (M_{\rm ej}/{\rm M}_\odot)^{5/6}$~yr. 
The transition time is given by $\tilde t_{\rm ST}=0.523$.

For a maximum residence time of the particles of $t_{\rm max} \sim R_{\rm s}/V_{\rm s}$, a simple approximation for the maximum energy for protons follows from Eq.~\ref{eq:accrate}:
\begin{eqnarray}
E_{\rm max}=\frac{3 Z e B V_{\rm s}R_{\rm s}}{\xi_\sigma c},
\end{eqnarray}
with $\xi_\sigma$ a relation between the compression ratio of the density and the magnetic field, 
where $\xi_\sigma=20$ for a shock where the magnetic field is parallel to the shock normal, and $\xi_\sigma=8$ for a shock with the magnetic field perpendicular to the shock normal.
The acceleration efficiency is highest around the Sedov-Taylor transition phase. 
For a simple ejecta profile and a homogeneous ISM ($n=0$, $s=0$, with $n$ the power law of the density in the ejecta envelope and $s$ the power law of the density of the surrounding medium) model, the corresponding typical radius of the blast wave and time of transition into this phase are given by:
\begin{eqnarray}
R_{\rm ST}& \approx & 2.23 \left(\frac{M_{\rm ej}}{{\rm M}_\odot}\right)^{1/3}n_0^{-1/3} {\rm pc}\\\nonumber
t_{\rm ST}& \approx & 209\ E_{51}^{-1/2}\left(\frac{M_{\rm ej}}{{\rm M}_\odot}\right)^{5/6}n_0^{-1/3} {\rm yr}.
\end{eqnarray}
At this point in the evolution of the SNR the maximum attainable energy for protons is:
\begin{eqnarray}
E_{\rm ST} \approx 107\ E_{51}^{1/2}\left(\frac{M_{\rm ej}}{{\rm M}_\odot}\right)^{-1/6} \left(\frac{B}{10 \mu G}\right) n_0^{-1/3}\; {\rm TeV},
\end{eqnarray}
with $n_0$ the number density of the surrounding medium, $E_{51}$ the explosion energy in units of $10^{51}$erg, 
and where the corresponding radius and age of the supernova remnant are taken from \citet{1995McKeeTruelove}. 
For electrons the maximum energy at sufficiently late times is limited by synchrotron losses and follows from the balance of the acceleration term 
and the synchrotron loss term in Eq.~\ref{eq:lossrate}. This occurs when the electron energy equals:
\begin{eqnarray}
\label{eq:Eemax}
E_{\rm sync}&=&\sqrt{\frac{18 \pi e}{\xi_\sigma \sigma_{\rm T} B}} \:  m_{\rm e} c V_{\rm s} \\\nonumber
& \approx & 11.6 \left(\frac{V}{6000\ {\rm km/s}}\right) \left(\frac{B}{10\ \mu G}\right)^{-1/2} {\rm TeV}, 
\end{eqnarray}
which corresponds to a synchrotron loss time-scale of \citep{2004vanderSwaluwAchterberg}:
\begin{eqnarray}
\label{eq:tsync}
t_{\rm sync} & \approx & 800 \left(\frac{B}{10\ \mu G}\right)^{-2} \left(\frac{E}{100\ {\rm TeV}}\right)^{-1} {\rm yr.} 
\end{eqnarray}

\begin{figure}
  \centering
 \includegraphics[width=0.5\textwidth]{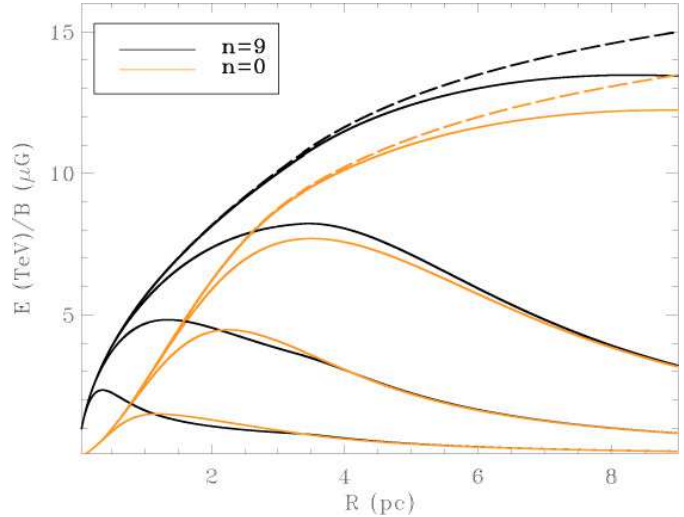}
 \caption[ ] {Maximum energy of relativistic electrons (solid) and protons (dashed) for different magnetic field strengths. 
The black lines corresponds to the $n=9$ model from \citet{1999TrueloveMcKee} and the coloured lines to the $n=0$ model. 
As long as the maximum energy is not limited by radiative losses, the influence of the shock velocity dominates the evolution of the maximum energy. 
The curves with the highest energies correspond to a magnetic field strength of $3\ \umu$G. 
The higher magnetic field strengths of $10$, $20$ and $50\ \umu$G limit the electron energy to subsequently lower values (normalized to the magnetic field strength).
    \label{fig:pmaxtimen0n9}}
  \end{figure}

In Fig.~\ref{fig:pmaxtimen0n9} we show the solution of Eq.~\ref{eq:lossrate} for the analytical estimate of the maximum energy as a function of the blast wave radius. 
From the equation we can see that the influence of the shock velocity on $p_{\rm max}$ is very strong. 
The different evolution of the shock velocity in the $n=0$ versus the $n=9$ model therefore is reflected in the strong differences of the maximum energy in the two models, 
as can be seen in Fig.~\ref{fig:pmaxtimen0n9}. 

\section{Method}
\label{sec:method}

\subsection{Stochastic differential equations}

We use the method of \citet{1992AchterbergKruells} to calculate the distribution function of cosmic rays. 
We describe the method below (details of the derivation for 2D spherical geometry can be found in Appendix~\ref{app:spherical}) and outline the set-up used in the hydro-simulation.

The propagation of relativistic particles through the plasma can be described by a random walk. To simulate scattering off MHD waves, such as Alfv\'en waves, 
the mean free path has a dependence on the particle momentum and the strength of the magnetic field. 
The presence of a magnetic field may cause the diffusion to be anisotropic, with diffusion along the field proceeding more rapidly than across the field. 
The acceleration and propagation of relativistic particles can be described by a phase space advection-diffusion equation \citep{1975Skilling, 1990Jones}:
\begin{eqnarray}
\frac{\partial F}{\partial t} = -\nabla_Z \cdot  \left({\bf U} F)+\nabla_Z \cdot ({\bf \kappa} \cdot \nabla_Z F \right) \; ,
\label{eq:advdiff}
\end{eqnarray}
where $F({\bf Z},t)$ is the particle distribution in phase space, $\nabla_{Z}$ is the gradient operator in phase space,
${\bf Z}\equiv ({\bf x},{\bf p})$ is the phase-space position vector, $\bf U = {\rm d} {\bf Z}/{\rm d}t$ is the phase space velocity and ${\bf \kappa}$ the diffusion tensor in phase space.

By reordering the operators we obtain a Fokker-Planck equation of the standard form;
\begin{eqnarray}
\label{eq:fokkerplanck}
\frac{\partial F({\bf Z},t)}{\partial t} = -\nabla_Z \cdot \left[{\bf \dot Z}F({\bf Z},t)-
\nabla_Z \cdot \left({\bf \kappa}F({\bf Z},t)\right)\right].
\end{eqnarray}
Here
\begin{equation}
            {\bf \dot Z} \equiv {\bf U} + \nabla_{Z} \cdot \kappa
\end{equation}
is an effective velocity that includes a drift term due to diffusivity gradients.
Equation~\ref{eq:fokkerplanck} is mathematically equivalent to a set of stochastic differential equations (SDEs) of the It\^o form \citep{1994Gardiner,1985Saslaw,1992AchterbergKruells}:
\begin{eqnarray}
\label{eq:ito}
d{\bf Z} = {\bf \dot Z} \: {\rm d}t +  \sqrt{2{\bf \kappa}} \cdot {\rm d} {\bf W} \; .
\end{eqnarray}
The SDEs contain, apart from a regular (deterministic) term $\propto {\rm d}t$, a stochastic term that models the random walk. The Wiener process ${{\rm d} \bf W}$ in the stochastic term satisfies $\langle dW_i \rangle=0$ and  $\langle dW_i dW_j \rangle=\delta_{ij}dt$.
The quantity $\sqrt{\kappa}={\bf T}$ represents a tensor ${\bf T}$ that should formally satisfy $T_{im}T_{mj}= \kappa_{ij}$.

A large number of statistically independent integrations of the SDEs creates a distribution of particles in phase space that corresponds to the solution of the corresponding Fokker-Planck equation \citep{1994Gardiner}. 

These SDEs can be used to calculate particle acceleration in a time-dependent flow, as first proposed in \citet{1992AchterbergKruells} 
and which has been applied  \citep[e.g.][]{1994KruellsAchterberg, 1999MarcowithKirk, 2004vanderSwaluwAchterberg, 2010MarcowithCasse} in various hydrodynamic codes. 
In 2D spherical geometry, the set of equations that needs to be solved to update the position of the particles in phase space (as we derive in Appendix~\ref{app:spherical}) is:
\begin{eqnarray}
{\rm d}u &=&-\frac{1}{3}({\bf \nabla \cdot V}) \: {\rm d}t - \lambda_s B^2 \sqrt{1+e^{2 u}} \: {\rm d}t \; ,  \nonumber \\
	&& \nonumber \\
{\rm d}R&=&\left( V_R+\frac{1}{R^2}\frac{\partial (R^2 \kappa)}{\partial R}\right) \: {\rm d}t + \sqrt{2 \kappa \: {\rm d}t}\: \xi_R \; , \\
	& & \nonumber \\
R \: {\rm d}\theta&=&\left( V_\theta+\frac{2 \kappa}{R \tan \theta}+\frac{1}{R}\frac{\partial \kappa}{\partial \theta}\right) \: {\rm d}t -\sqrt{2 \kappa \: {\rm d}t}\: \xi_\mu \; , \nonumber
\end{eqnarray}
where $V$ is the plasma velocity, $u = \ln(p/mc)$  and $\mu=\cos\theta$. 
The stochastic term now contains a random number $\xi_i$ drawn from a unitary normal distribution such that its statistical average satisfies $\langle \xi_i \rangle=0$ and $\langle \xi_i \xi_j \rangle=\delta_{ij}$, where $i,j$ stand for $R$ and $\mu$. Note that these equations solve for $\tilde F=R^2 F$ rather than for $F$. 

In 1D slab geometry with flow velocity $V(x \: , \: t)$ these equations simplify to:
\begin{eqnarray}
\label{1DslabSDE}
{\rm d}u&=&-\frac{1}{3} \left( \frac{\partial V}{\partial x} \right) \:  {\rm d}t - \lambda_s B^2 \sqrt{1+e^{2 u}} \: {\rm d}t \; , \nonumber \\
		& & \\
{\rm d}x&=&\left( V+\frac{\partial \kappa}{\partial x}\right) {\rm d}t +\sqrt{2 \kappa \: {\rm d}t} \: \xi_x\; . \nonumber
\end{eqnarray}

Some caution should be exercised when translating these equations into numerical schemes in the case that large magnetic field gradients are present.
When Bohm diffusion is assumed, gradients in the magnetic field strength induce gradients in the diffusion coefficient. 
This creates the following problem: In Eq.~\ref{eq:Roperator} we see that the divergence of the diffusion tensor shows up in the effective velocity of the scattering centre of our cosmic rays. 
In simple schemes the strong gradient in the magnetic field causes the particles to under-sample the shock statistically, giving rise to a momentum spectrum of the accelerated particles that is steeper than can be expected in reality. 
A more sophisticated numerical scheme is needed to make sure that the spectrum reflects the physics, rather than mathematical artifacts (Achterberg \& Schure, in preparation).  Its implementation will be presented in a follow-up paper. 
Here we will only consider parallel shocks where the magnetic field strength is constant across the shock.

The resulting power-law spectrum should approach the analytical solution of a strong shock, 
where the compression ratio depends only on the specific heat ratio $\gamma$ of the plasma, and the slope of the distribution depends only on $r$. 
The compression ratio is given by $r=\rho_2/\rho_1=(\gamma+1)/(\gamma-1)$, where $\rho$ denotes the plasma density. 
For a value of $\gamma=5/3$, $r=4$, whereas for $\gamma=1.1$, $r=21$. In terms of $u = \ln(p/mc)$ one has:
\begin{eqnarray}
F(u) = pF(p) \propto p^{-(q-1)},
\end{eqnarray}
with $q$ given by Eq.~\ref{eq:q}.

\subsection{Cosmic ray injection}
\label{sec:injection}

As mentioned in \S~\ref{sec:intro}, the injection of cosmic rays at the shock front is still a poorly understood process. 
Observations hint at efficient injection, particularly for parallel shocks \citep{2008BerezhkoVoelk}, 
but it remains problematic to accelerate electrons from the thermal pool to sufficiently high energies for them to be picked up by the DSA process. 
Since we can only model the highest-energy end of the particle distribution, we leave this problem to others (see \citet{2009SironiSpitkovsky} for relativistic shocks), 
and assume that the rate of particle injection is proportional to the mass that is swept up per unit time by the blast wave. 
Since we treat the cosmic rays in the test-particle limit, 
the exact number that is injected is not so important, but the relative number of injected cosmic rays 
with time influences the momentum distribution and the average acceleration time of the cosmic rays. 

For a homogeneous ISM, the number of particles that is injected as the shock expands a distance ${\rm d}R$ in radius scales as: 
\begin{eqnarray}
{\rm d}N(p)\propto R^2 \: {\rm d}R \: \delta(p-p_0),
\end{eqnarray}
We assume mono-energetic injection with $p_0$ the injection momentum, which for Bohm diffusion is restricted for numerical reasons, as will be explained below (Eq.~\ref{eq:pinj}).
For a CSM that is shaped by the stellar wind of the progenitor the density scales with $\rho \propto R^{-2}$. 
In this case the injection proceeds uniformly with respect to the blast wave radius:
\begin{eqnarray}
{\rm d}N(p)\propto {\rm d}R \:  \delta(p-p_0).
\end{eqnarray}

\subsubsection{Minimum injection momentum}

In our models, the particles are injected with relativistic energies. 
They need to be able to cross the shock in one scattering mean free path, otherwise they will be adiabatically accelerated and we will not see the desired effect of DSA. 
The particles are all injected with the same (relativistic) energy and the power law naturally follows provided the steps used for the integration of the SDEs satisfy
\begin{eqnarray}
\label{eq:deltaxcriterion}
\Delta x_{\rm adv} \simeq V_{s} \: \Delta t_{\rm dsa} \ll \Delta x_{\rm s} \ll \Delta x_{\rm diff} = \sqrt{2 \kappa \: \Delta t_{\rm dsa}}.
\end{eqnarray}
Here $\Delta t_{\rm dsa}$ is the time step used to integrate the SDEs, and $\Delta x_{\rm s}$ is the width of the shock transition, typically a few grid cells.
In the case of Bohm diffusion, the particles are inserted with an energy that makes it possible to meet the criterion in Eq.~\ref{eq:deltaxcriterion} 
for the simulation of DSA from the earliest injection time ($t_0=31$~yr) onwards. 
This puts the following constraint on the diffusion coefficient: 

\begin{eqnarray}
\kappa \gg \frac{\Delta x_{\rm s}^2}{2 \Delta t_{\rm dsa}} \; ,\label{eq:xdiff}
\end{eqnarray}
with 
\begin{eqnarray}
\kappa=\frac{r_g c}{3}=\frac{p c^2}{3 Z e B_{\rm max}} \quad {\rm (cgs)},\label{eq:bohm}
\end{eqnarray}
where $Z$ is the charge number of the particle, which we take to be $1$ since we only consider protons and electrons.
The requirement of the advective step, $\Delta x_{\rm adv} < \Delta x_{\rm s}$, in practise translates to a time-step requirement \begin{eqnarray}
\Delta t_{\rm dsa} &< &\frac{\Delta x_{\rm s}}{v_r}. 
\label{eq:dtdsa}
\end{eqnarray}
Since the shock thickness is typically resolved in about 5 grid cells, the time step for calculating the diffusion can be larger than that used for the MHD time step. 
This one is limited by the courant condition where we normally use $\Delta t_{\rm hydro}=0.8 \Delta x_{\rm grid}/v$. 
We therefore `supercycle' (i.e.~increase the timestep) the diffusion of the particles to save computational time, and calculate the appropriate time step $\Delta t_{\rm dsa}$
for diffusion throughout the simulation.

The constraint on the timestep can be used to evaluate the valid range for the diffusion coefficient by combining Eqns.~\ref{eq:xdiff} to \ref{eq:dtdsa}. 
The minimum injection energy for this model is then given by:
\begin{eqnarray}
\label{eq:pinj}
p_{\rm inj} = \frac{3 q B \kappa}{c^2} > \frac{3 q B v_{\rm max}^2 \Delta t_{\rm dsa}}{2 c^2}. 
\end{eqnarray}

The scaling of the injection energy with magnetic field also means that our resolution requirement depends on the magnetic field strength assumed in the simulation. 
We want the injection energy to be at least as low as $\sim 1$~TeV, such that it is well below the maximum energy as can be roughly determined by Eq.~\ref{eq:Eemax}. 
At lower energies, the spectrum inside the SNR will be a featureless power law as the scattering mean free path of the particles is much less than, say, 
the radius of curvature of the shock, and losses are unimportant. 
The maximum number of particles injected at the end of the simulation is $10^6$. 
Particle splitting is applied when the particle energy increases with a factor of $e$, in order to reduce Poisson noise in the distribution at high energies.

\subsection{The hydrodynamics}
\label{sec:ejecta}

We have incorporated the equations for calculating diffusive shock acceleration into the AMRVAC framework. 
This allows us to solve the particle spectrum concurrently with the (magneto-)hydrodynamics of the flow, in this case the evolution of the supernova remnant. 
We model the system in 1D spherical coordinates. We use a radial range of $3.0\times10^{19}$~cm ($\sim 10$~pc) and a base grid of 180 cells. 
Refinement of a factor two is applied dynamically where strong density and velocity gradients are present, up to 7 to 10 levels. The effective resolution is then $46080-184320$~cells, or $1.6\times10^{14}-2.6\times10^{15}$~cm. This is sufficient for our simulations of the system with $B$ up to $20\ \umu$G. 
The Euler equations are solved conservatively using a TVDLF scheme with minmod limiting on the primitive variables \citep{1996TothOdstrcil}. 
Although this particular scheme does not resolve the shock within the least possible number of grid cells, the scheme is robust and does not have problems in dealing with the initial conditions for the supernova ejecta, such as large density gradients.

The supernova ejecta are inserted in the inner $0.1$~pc of the grid. The density and velocity profile are modelled according to the self-similar solution of \citet{1984Chevalier, 1999TrueloveMcKee}. The velocity increases linearly to the outer edge of the ejecta, whereas the density profile consists of a constant density core and a powerlaw envelope with a slope of $n=9$, interpolated as:
\begin{eqnarray}
\rho_{\rm ej}(r) =\frac{\rho_{\rm core}}{1 + (r/r_{\rm core})^{n} }.
\end{eqnarray}
The core radius and central density are such that the mass and energy of the ejecta are respectively $M_{\rm ej}=3.5$~M$_\odot$ and $E_{\rm ej}=10^{51}$~erg. 

\begin{figure}
   \centering
   \includegraphics[width=0.5\textwidth]{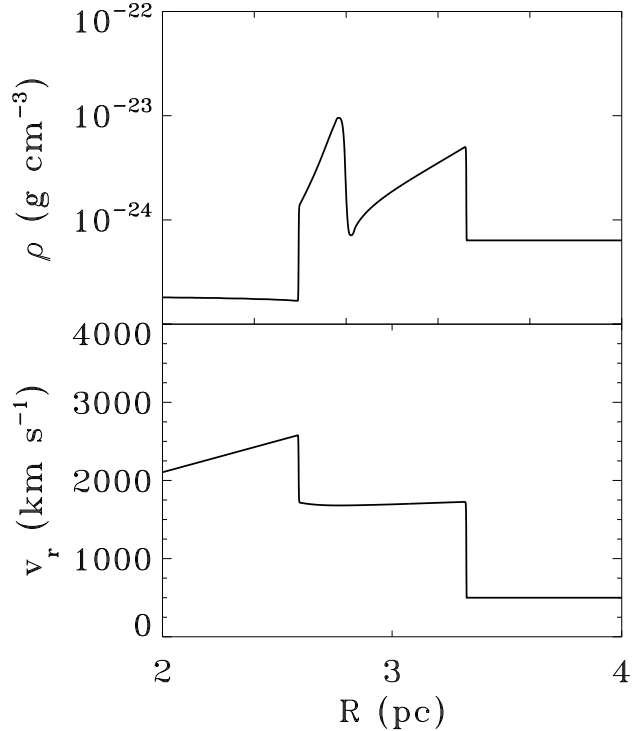}
      \caption[ ] {
      Density and velocity profile of supernova ejecta at a time $t=600$~yr after explosion into a homogeneous medium. The shown profiles are for a plasma with an adiabatic index of $\gamma=5/3$, yielding the expected compression ratio $r=4$ at the shocks.
      \label{fig:rhovism}}
   \end{figure}

\begin{figure}
   \centering
   \includegraphics[width=0.5\textwidth]{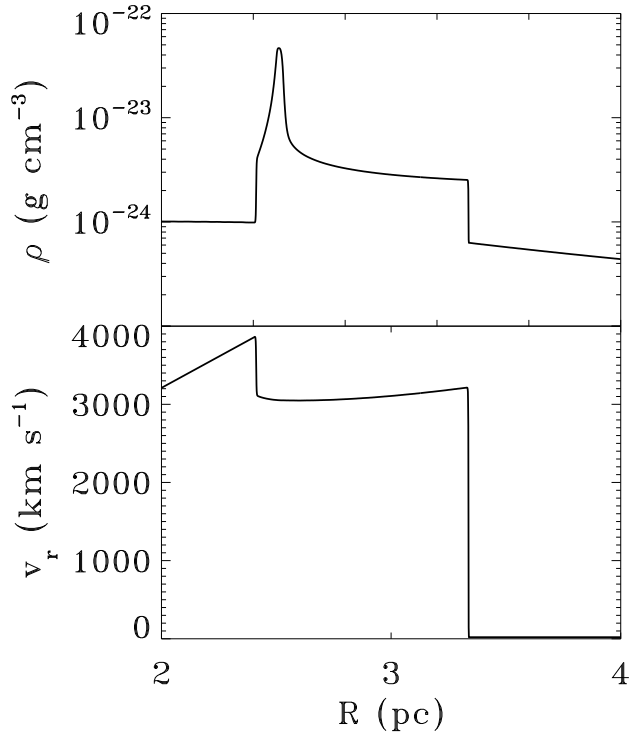}
      \caption[ ] {
      Density and velocity profile of supernova ejecta at a time $t=600$~yr after explosion into a $\rho \propto R^{-2}$ medium. Compared to the evolution of a SNR in a homogeneous ISM, the reverse shock is much farther inwards and the distance between the contact discontinuity and the forward shock much larger.
      \label{fig:rhovcsm}}
   \end{figure}

In Fig.~\ref{fig:rhovism} we show the density and velocity profile for a SNR, $600$~yr after explosion into a homogeneous medium. In this simple one-dimensional model four different regions can be identified: From inside out first we encounter the freely expanding ejecta, separated from the shocked ejecta by the reverse shock. Outside the contact discontinuity is the compressed ISM and the blast wave that separates the SNR from the undisturbed ISM. In Fig.~\ref{fig:rhovcsm} we show the same for the supernova remnant in a CSM. The evolution of the SNR is substantially different in the two cases. The separation between the blast wave and the ejecta is larger in the CSM model. This is due to the deceleration of the blast wave, which initially is high for the CSM model, but soon becomes slower than for the ISM background, as can also be seen in Fig.~\ref{fig:rvshocktimemulti}, which will be discussed later in Sect.~\ref{sec:csmism}.

\section{Test models}
\label{sec:testmodels}
In this section we will present the results from some test cases, in order to compare the numerical results with the analytical ones, 
and to determine the dependence of the spectrum on the chosen geometry.

\subsection{Analytical shock profile}
We will first present the results from a calculation where we describe the shock using hypertangent profile in plane parallel geometry.
In the shock's reference frame the velocity $V(x)$ along the shock normal is
\begin{eqnarray}
V(x)&=&\left(\frac{r+1}{2r}\right)-\left(\frac{r-1}{2r}\right)\tanh\left(\frac{x-x_{\rm s}}{\Delta x_{\rm s}}\right) \; .\label{eq:tanh}
\end{eqnarray}
The shock has a thickness $\Delta x_{\rm s}$ (in our case: $3.75\times10^{-2}$ in arbitrary units) and is located at a location of $x=x_{\rm s}$ (in our case at $x=9$).
The shock compression ratio is chosen to be $r = 4$, the value for a strong shock in a $\gamma = 5/3$ gas. The time step is chosen to satisfy the condition in Eq.~\ref{eq:deltaxcriterion}: $\Delta t= 1.35\times 10^{-2}$ and the shock velocity is normalized to $V_{\rm s}=1$.
The diffusion coefficient is constant: $\kappa=0.28$. In this case, the numerically obtained cosmic ray distribution should closely match the analytical one. 

\begin{figure}
  \centering
 \includegraphics[width=0.5\textwidth]{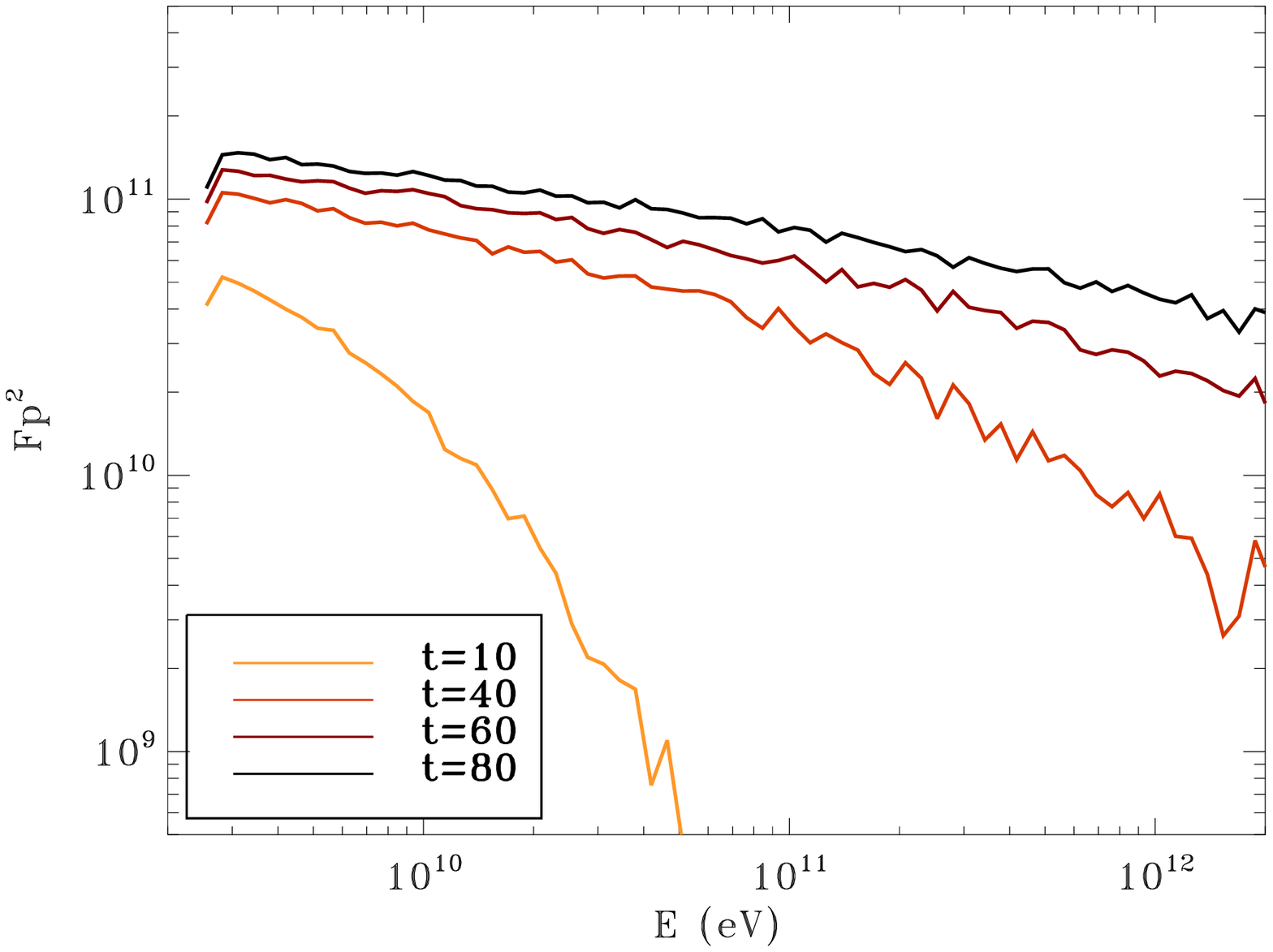}
 \includegraphics[width=0.5\textwidth]{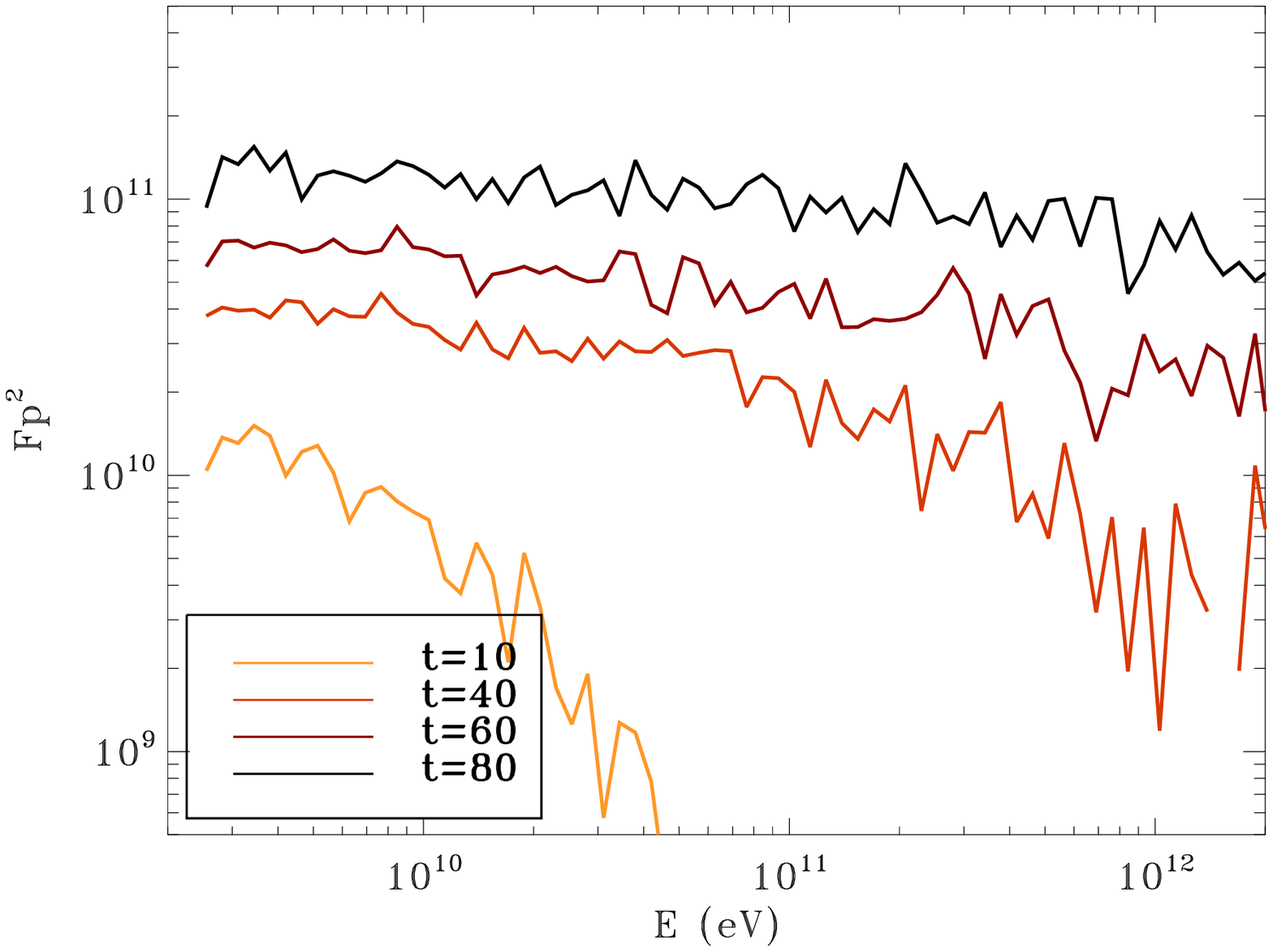}
 \caption[ ] {Spectrum as a function of time for accelerated particles in slab geometry and with a hypertangent velocity profile. The top panel shows the total spectrum, the bottom panel the spectrum at the location of the shock.
     \label{fig:tanhFptime}}
  \end{figure}

In Fig.~\ref{fig:tanhFptime} we show the spectrum of the particles that are located near the shock, and that of all the particles. The number of particles that is introduced at the shock increases linearly with time, and they are all introduced with the same energy. The maximum number of particles used in these simulations is $10^6$. The time evolution shows that it takes time before the numerical slope approaches the analytical steady-state result, which in our figures would correspond to a horizontal line (meaning $q=2$ for our chosen compression ratio of $r=4$). While the spectrum at the shock indeed approximates the expected result, the spectrum of the whole particle population is steeper. 
This is to be expected, because the overall spectrum consists of a superposition of spectra with different maximum acceleration times (for which Fig.~\ref{fig:tanhFptime}, lower panel, gives a sample), and hence different cut-off energies. The overall spectrum is therefore steeper than the spectrum at the shock front.

\subsection{Effects of the shock geometry: planar and spherical}
In this test model we explore the difference on the slope of the energy spectrum of the particles that arises as a result of the adopted geometry. 
We will compare this to the analytical model, which assumes a  steady state solution for a plane parallel shock.
We set up the supernova ejecta in a homogeneous density medium and let it evolve for 1500~yr. We artificially induce slab or spherical geometry, taking into account the volume- and surface elements depending on the chosen geometry, such that adiabatic losses are properly accounted for.
In a SNR the radius of curvature is much larger than the typical diffusion path of a particle, and slab geometry therefore is normally considered to be a good approximation. 
However, we find that there are differences in the slope of the spectrum when spherical geometry is taken into account, and argue that geometry cannot simply be neglected.  

\begin{figure}
  \centering
 \includegraphics[width=0.5\textwidth]{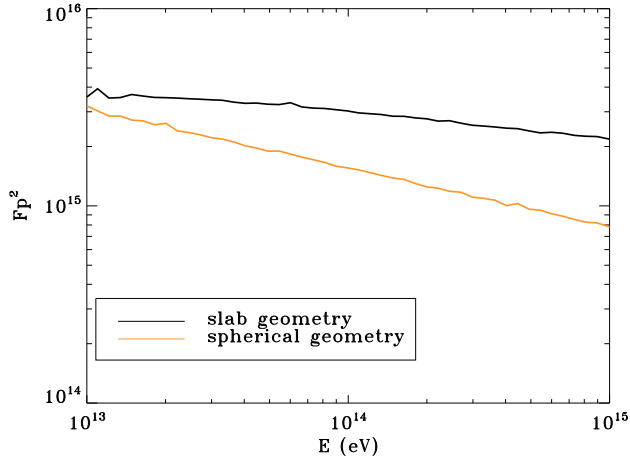}
 \caption[ ] {Proton spectrum for the case of a constant diffusion coefficient. The black line shows the spectrum for a shock in plane parallel geometry. 
The coloured line shows the case for spherical geometry. The maximum energy extends up to unrealistically high energies for which physically the particles would not be confined to the supernova remnant.
    \label{fig:slabsphereFpshtime}}
  \end{figure}

Figure~\ref{fig:slabsphereFpshtime} shows the differences in the energy spectrum that arise 
due to the choice of geometry. In plane-parallel geometry the spectrum is closer to the analytical test case as presented in the previous section. 
In spherical geometry, the spectrum is somewhat steeper, with a slope $q\approx2.15$. This is to be expected since particles downstream of the shock experience adiabatic losses that are proportional to $V/r$. This reduces the mean energy gain they experience at the shock, which steepens the spectral slope. 
Additionally, the exact slope in a time-dependent calculation depends on the injection rate and its dependence on time.
  
\section{Results}
\label{sec:snr}

\subsection{Influence of the choice of diffusion coefficient}
\label{sec:kappa}
The particle spectrum that results from the diffusive shock acceleration process in supernova remnants depends on a number of factors as described by the equations in the previous sections. 
In this section we show how the assumption for the diffusion coefficient leaves its imprint on the slope of the spectrum.

\subsubsection{Constant diffusivity}

\begin{figure}
  \centering
 \includegraphics[width=0.5\textwidth]{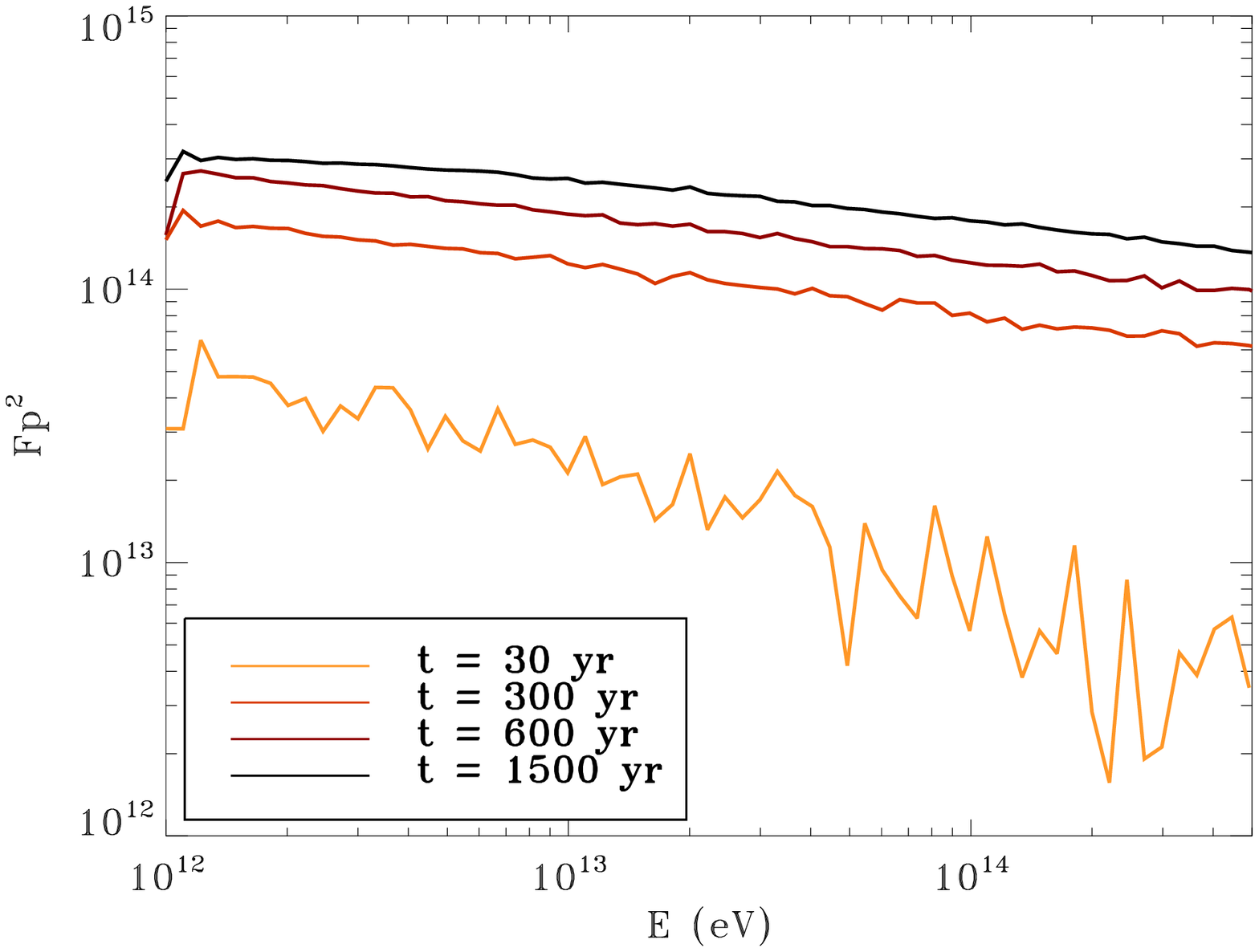}
 \includegraphics[width=0.5\textwidth]{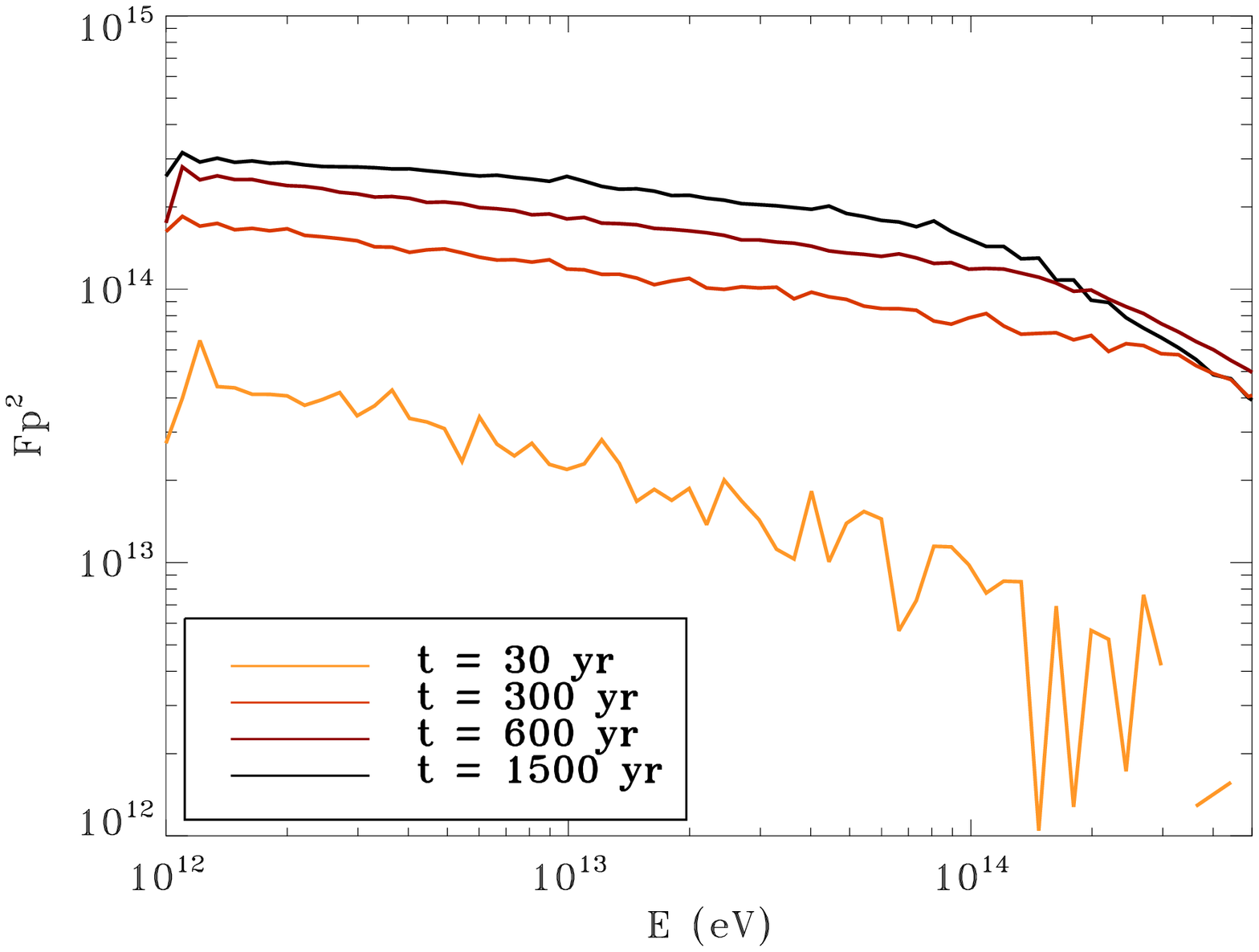}
    \caption[ ] {Spectrum as a function of time for relativistic protons (top) and electrons (bottom) for a SNR in a constant density ISM. The maximum energy of the electrons is limited by synchrotron losses for a field of $B=10\ \umu$G. The diffusion coefficient is a constant. The top curves are for a SNR age of: $t\approx 1500$~yr. The subsequent lower and lighter coloured curves are for times: $t\approx 300$, $150$, and $30$~yr.
    \label{fig:kconstFptime}}
  \end{figure}

First, we consider the case of proton and electron acceleration with a constant diffusion coefficient at a spherical shock that expands into a constant density ISM. 
We fix $\kappa$ to the value for which the particles satisfy criterion (\ref{eq:deltaxcriterion}) for the diffusion length. 
In Fig.~\ref{fig:kconstFptime} we show the integrated distribution of protons and electrons as function of energy, represented as $p^2 \: F(p)$, for different times. 
At late times, the solution well below the cut-off energy should approach the analytical solution (corresponding to a horizontal line in this representation). The reason why the slope remains steeper is because spherical geometry is taken into account, something that is not considered for the analytical solution. As found in the previous section, the spectral slope of cosmic rays in spherical geometry approximates 2.15 rather than 2. The same slope is found here.

For protons, the typical acceleration time-scale now only depends on the shock velocity (Eq.~\ref{eq:accrate}). 
For electrons, the maximum energy is limited by the balance between acceleration rate and loss rate. 
The acceleration rate decreases with the shock velocity as $V_{\rm s}^2$ and, hence, with time. The loss rate increases with energy. 
The effect is therefore strongest at late times, which can be seen in Fig.~\ref{fig:kconstFptime}, where the late-time curves have the maximum electron energy at a lower value than the early-time curves.

\subsubsection{Bohm diffusion}

\begin{figure}
  \centering
 \includegraphics[width=0.5\textwidth]{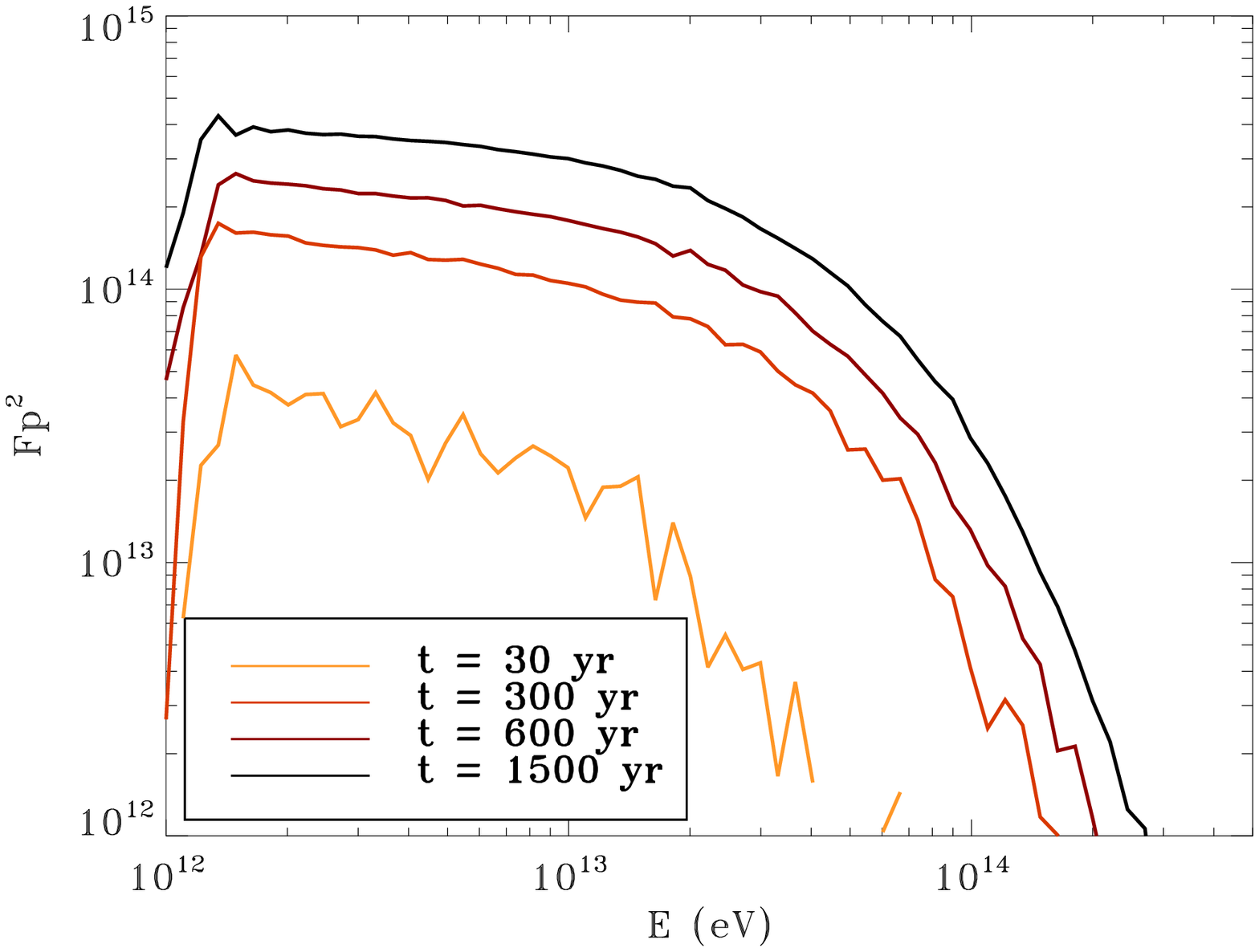}
 \includegraphics[width=0.5\textwidth]{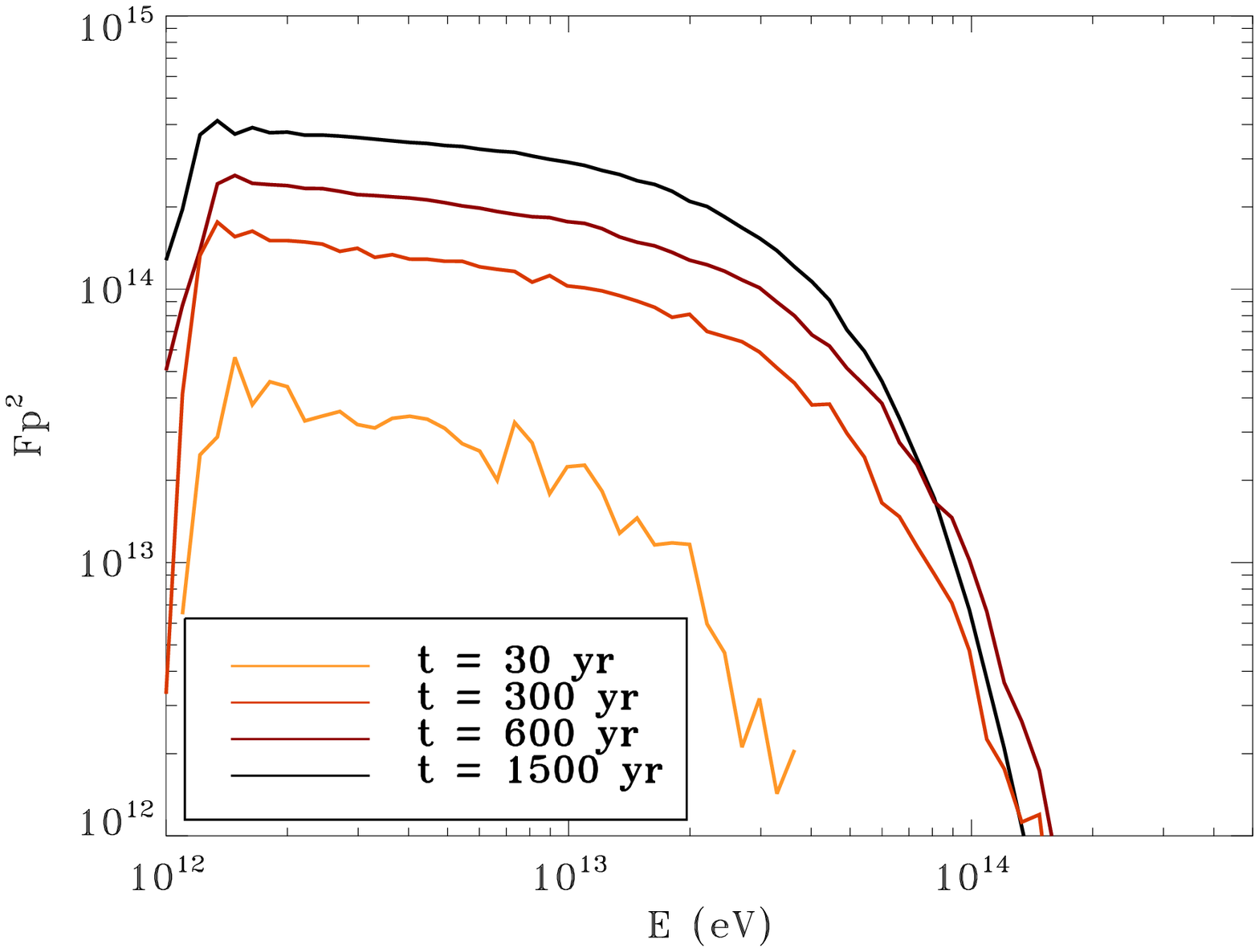}
 \caption[ ] {Spectrum as a function of time for relativistic protons (top) and electrons (bottom) for a SNR in a ISM. 
The maximum energy of the electrons is limited by synchrotron losses for a field of $B=10\ \umu$G. The approximation of Bohm diffusion is used for the scattering of the particles. 
As can be seen from the abrupt cut-off at the low-energy end of the spectrum, the injection energy of the particles is around $1$~TeV.
    \label{fig:ismn10Fptime}}
  \end{figure}

When Bohm diffusion with $\kappa \propto E$ is assumed, the acceleration time-scale increases, for a given shock velocity, with particle energy as $t_{\rm acc} \propto E$ (Eq.~\ref{eq:tacc}). 
It therefore takes longer for the higher energy particles to assume the equilibrium power law slope predicted by steady-state calculations. 
As can be seen in Fig.~\ref{fig:ismn10Fptime} this is visible in the spectrum as a smooth roll-over of the spectrum beyond a certain energy. 
For sufficiently `old' electrons synchrotron losses lower the maximum energy and lead to a steeper cut-off than for protons.

The different curves show the spectrum for different ages of the SNR. While for protons the cut-off energy steadily increases with time, this is not the case for the electrons. 
As long as the electron spectrum is determined by the time available for acceleration (the age of the system), it closely resembles the proton spectrum. 
However, once the electrons reach an energy where the synchrotron loss time-scale becomes comparable to the acceleration time-scale and/or the age of the system, 
the spectrum  deviates more and more from the proton spectrum, resulting in a lower cut-off energy and a steeper roll-over.

The spectral slope at the lower-energy end of the electron and proton spectrum is the same and does not depend on the choice of diffusion coefficient. 
It equals $q\approx 2.15$, the same value we have found in the test-case of the hypertangent shock profile in spherical geometry. 
This is to be expected, since the injection rate in both the test-case and the supernova remnant ISM models is the same, 
and adiabatic losses operate in the same manner. 

In the case of a constant diffusion coefficient, $\kappa$ was set to the value equal to the 
Bohm diffusion coefficient for the lowest energy particles (at $p=p_0$). 
The acceleration time for the high-energy particles is therefore relatively short compared 
to the $\kappa_{\rm B}$ models, moving the cut-off to higher energies. 

\subsection{Choice of equation of state}

\begin{figure}
  \centering
  \includegraphics[width=0.5\textwidth]{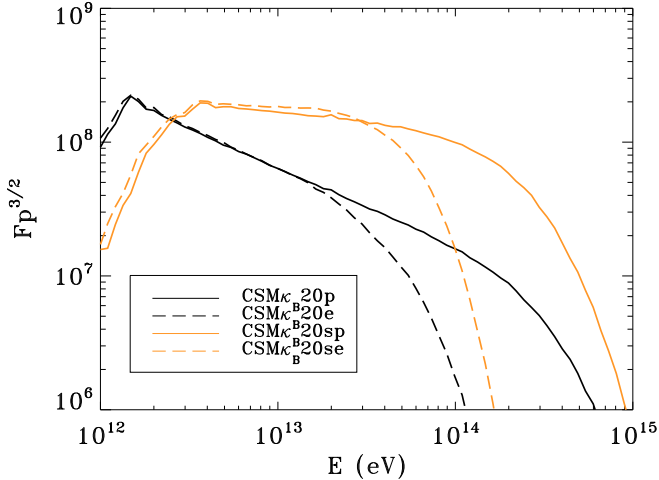}
 \caption[ ] {The proton (solid line) and electron (dashed line) spectra as resulting from a SNR evolving into a CSM with a softer equation of state: the CSM$\kappa_B20$s model (coloured lines). The maximum energy of the electrons is limited by synchrotron losses because of the high magnetic field strength ($B=20\ \umu$G). The black lines indicate the spectra for the same model but with $\gamma=5/3$ 
    \label{fig:eosFp49}}
  \end{figure}

In Fig.~\ref{fig:eosFp49} we show the energy spectrum for protons and electrons accelerated at a shock in a gas with adiabatic index $\gamma=4/3$. 
The shock propagates into a CSM with a $\rho \propto R^{-2}$ density profile.
We represent the spectrum as $p^{3/2} \: F(p)$ and compare the results to that of a $\gamma=5/3$ plasma model. 
For $\gamma=4/3$ the expected spectral slope at a steady shock is $q=1.5$ (see Eq.~\ref{eq:q}). 
We see that the simulations bear out this expectation. 
The other model parameters are $B=20\ \umu$G, $t_{\rm max}\approx 1500$~yr, and a CSM background. 
For a shock propagating into a constant-density ISM the same change in the spectral slope will result.

Varying $\gamma$ is a nice test to see if the slope changes according to expectations, but also is a way of determining in which direction results change when 
cosmic ray acceleration is efficient enough to change the equation of state. 
In reality, the cosmic ray pressure that causes the softening of the equation of state is localized in the  region around the shock, where most of the relativistic particles are located. 
We do not include this position dependence and keep the adiabatic index fixed to the same value over the entire grid.

\begin{figure}
  \centering
  \includegraphics[width=0.5\textwidth]{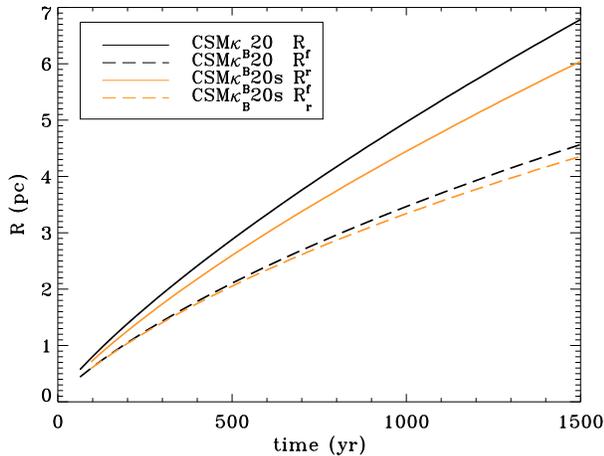}
 \caption[ ] {The radius of the blast wave (solid line) and the reverse shock (dashed line) is plotted for the CSM model with $\gamma=4/3$ (coloured lines) and compared to the CSM $\gamma=5/3$ model (black). The higher compression ratio of the $\gamma=4/3$ model leads to the shorter distance between the blast wave and the reverse shock. The blast wave radius is typically smaller and thus its velocity, too, resulting in a lower $E_{\rm max}$ for the protons accelerated in this simulation. 
    \label{fig:eosshocktime}}
  \end{figure}

In our approach the softer equation of state results in a higher compression at the shock and a slower blast wave, as can be seen in Fig.~\ref{fig:eosshocktime}. 
The cut-off energy for protons therefore lies at a lower value than in the $\gamma=5/3$ model. 
Since we assume a fairly high magnetic field in this simulation and the $\gamma = 5/3$ case we use for comparison, the electron energy is limited by synchrotron losses. 
Therefore in both models electrons have a similar value for $E_{\rm max}$.

\subsection{CSM versus ISM models}
\label{sec:csmism}

In this section we compare the results from the CSM and ISM models with $B=3\ \umu$G. Bohm diffusion is assumed in both cases.
As we already saw in the analytical calculations in Sect.~\ref{sec:time-scalesanalytical}, 
the maximum particle energy depends strongly on the shock velocity $V_{\rm s}$. 
The density of the medium in which the SNR expands affects the shock velocity and therefore leads to differences in the cosmic-ray acceleration rate between the CSM and the ISM models.

\begin{figure}
  \centering
 \includegraphics[width=0.5\textwidth]{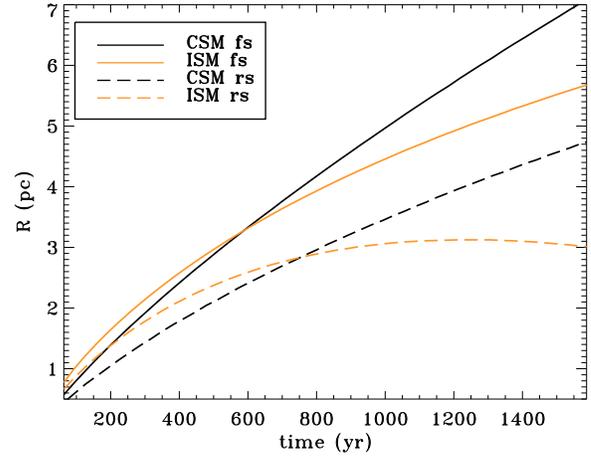}
 \includegraphics[width=0.5\textwidth]{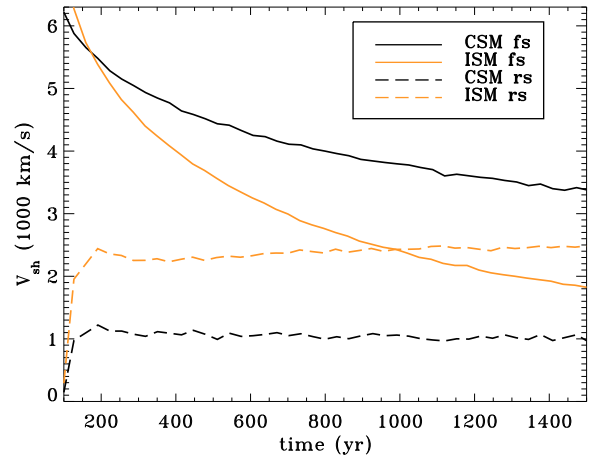}
 \caption[ ] {Top: evolution of the radius of the forward (solid line) and the reverse (dashed line) shock of the SNR in the CSM (black) and ISM (yellow) models. Bottom: the corresponding velocity of the blast wave and the reverse shock (relative to the upstream velocity). 
    \label{fig:rvshocktimemulti}}
  \end{figure}

A blast wave expanding into the CSM hits a relatively dense medium early on in its evolution. As a result, the initial velocity is smaller but the deceleration proceeds more slowly,
as the swept-up mass increases with radius as $M_{\rm sw} \propto R$, as opposed to $M_{\rm sw} \propto R^3$ in the ISM case.
Ultimately this results in a shock with a higher velocity at the end of the simulation ($1500$~yr). 
In the ISM model the initial shock velocity is higher, but the deceleration much more severe, and the maximum attainable particle energy is lower, at
least for the model parameters used here. 
In Fig.~\ref{fig:rvshocktimemulti} we show the evolution of the radius of the forward and the reverse shock in the top panel, 
and in the bottom panel the velocity of the blast wave for the SNR in CSM, ISM.

\begin{figure}
  \centering
 \includegraphics[width=0.5\textwidth]{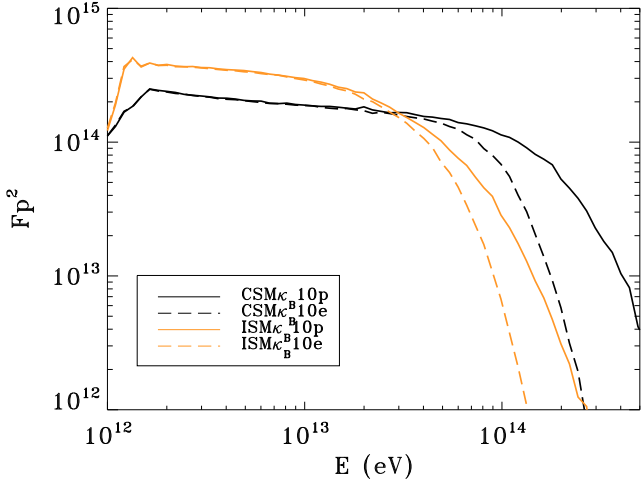}
 \caption[ ] {Spectra for protons (solid) and electrons (dashed) for CSM (black) and ISM (coloured) models with $B=10\ \umu$G. 
    \label{fig:ismcsmFptime}}
  \end{figure}

The injection rate of particles is taken to be proportional to the mass that is swept up per unit time 
by the blast wave. 
As a result the age distribution of cosmic rays in the CSM and ISM models differs. 
This difference affects both the maximum energy and the shape of the overall spectrum. 
This is shown in Fig.~\ref{fig:ismcsmFptime}. The proton/electron spectrum in the CSM model, represented as $p^2 \: F(p)$, is slightly concave. 
This arises because of the higher fraction of `old' particles in the CSM model compared to the ISM model. On average, these older particles have a higher energy and
(for Bohm diffusion) a larger diffusivity. Low-energy particles on the other hand are more rapidly swept away from the shock. This simulation shows the differences
that arise in a time-dependent calculation with respect to the results at a steady (unchanging) shock. 

\begin{figure}
  \centering
 \includegraphics[width=0.5\textwidth]{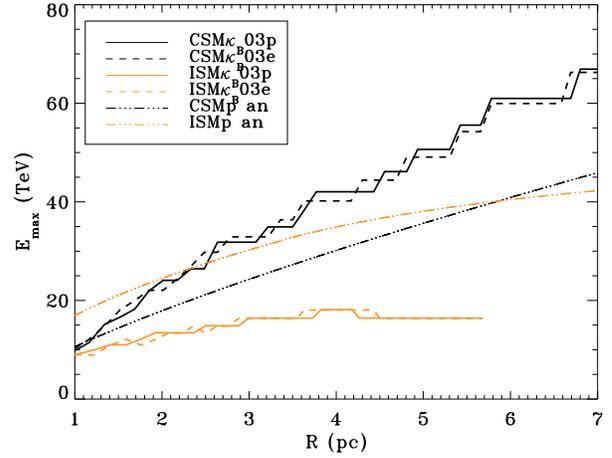}
 \caption[ ] {$E_{\rm max}$ for protons (solid) and electrons (dashed) for CSM (black) and ISM (coloured) models with $\kappa_B$ and $B=3\ \umu$G. The analytical solution for the CSM (ISM) model (without radiation losses) is plotted with dash-dot-dots in black (coloured). 
    \label{fig:pmaxismcsm}}
  \end{figure}

In Fig.~\ref{fig:pmaxismcsm} we show the maximum energy of the particles as extracted from the simulations.
Since the slope of the overall spectrum in this case is about $q=2.15$, we define $E_{\rm max}$ as the e-folding energy, 
where $p^{2.15} \: F(p)$ for the cumulative spectrum decreases to a value smaller than $1/e$ times the value at lower energies. 
The higher average velocity of the blast wave when it evolves into a CSM increases $E_{\rm max}$ by a factor $2-4$ for the CSM models. 
Due to the low magnetic field strength, the synchrotron loss time for this model is significantly longer than the running time of the simulations ($\sim 4\times 10^4$~yr versus $1500$~yr). 
Therefore there is no significant difference between the proton and the electron spectrum. In figure  \ref{fig:pmaxismcsm} we overplotted the maximum energy as calculated analytically in Sect.~\ref{sec:method} for protons (since, because of the low magnetic field strength in this particular model that for electrons is about the same). 

There is a clear difference between $E_{\rm max}$ derived from the analytical estimate with that from the simulations. For the CSM, the analytical model underestimates this value, 
whereas for the ISM the analytical estimate is significantly larger than the value we derive from the simulations. 
This conclusion holds when we calculate $E_{\rm max}$ as the e-folding energy for the spectrum at the shock, but also for the spectrum of all particles in the SNR. 
We attribute this to the different age distributions in the CSM and ISM models, as explained above.

\subsection{Results for different magnetic field strengths}
\label{sec:magneticfield}

In the previous section we showed results for a magnetic field strength of $3\ \umu$G. 
While this value is too low to induce significant differences between proton and electron spectra in these models, differences arise for stronger magnetic fields.
The maximum proton energy increases as $E_{\rm max} \propto B$ in the radiation loss-free case that applies here.The electron energy decreases roughly as $E_{\rm max} \propto 1/\sqrt{B}$ for electrons if synchrotron losses are important. 

\begin{figure}
  \centering
 \includegraphics[width=0.5\textwidth]{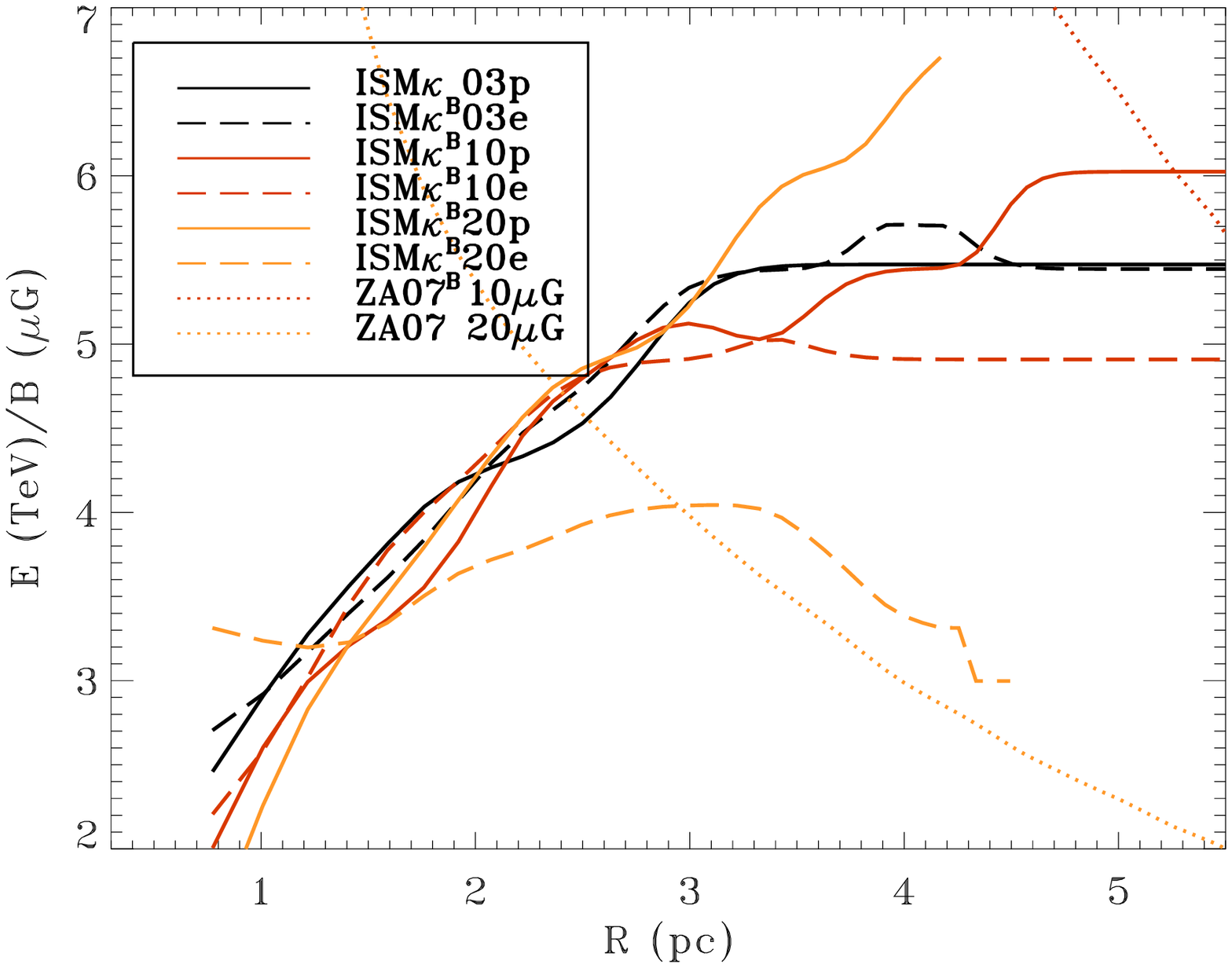}
 \includegraphics[width=0.5\textwidth]{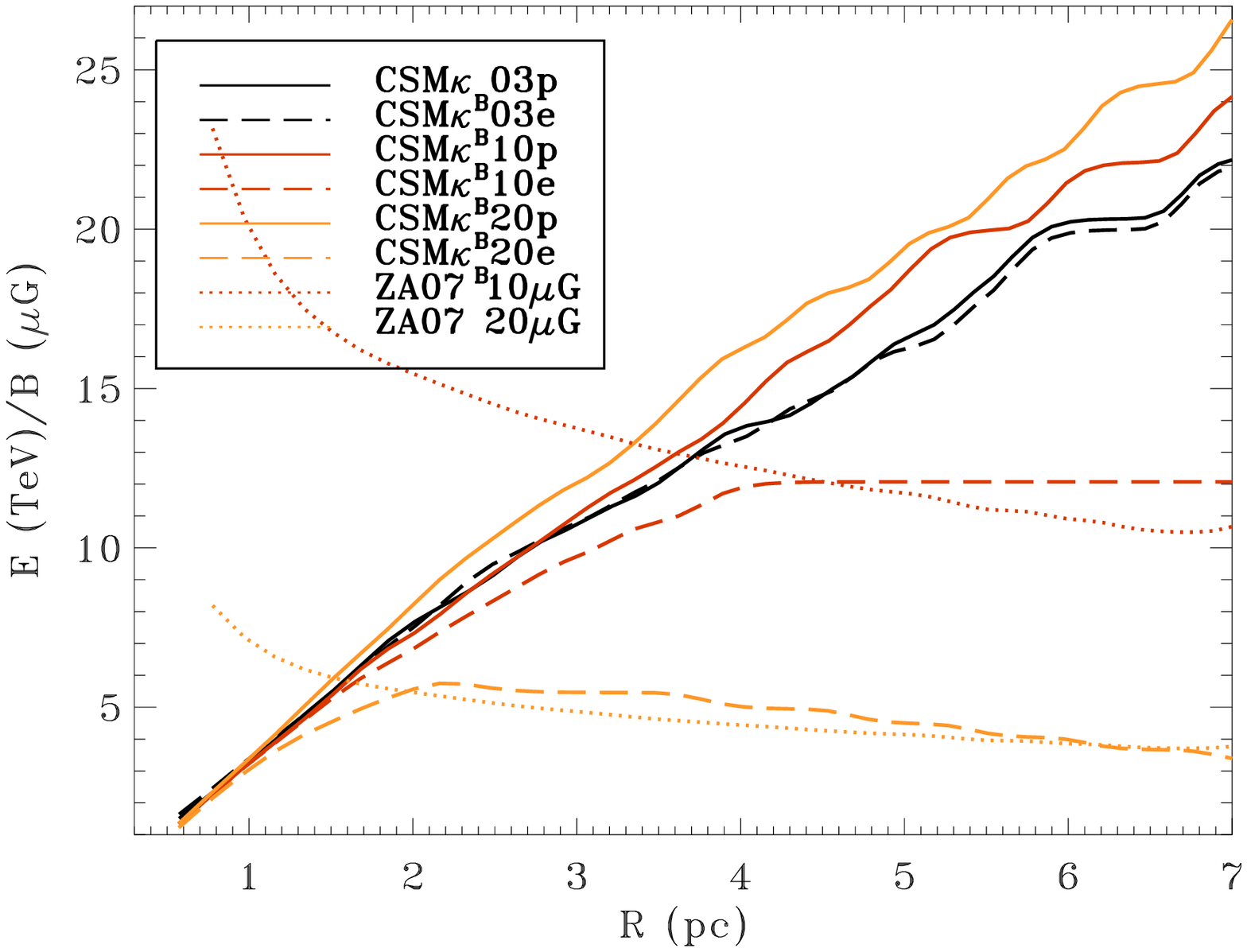}
 \caption[ ] {Maximum energy as a function of blast wave radius for relativistic electrons (dashed) and protons (solid). The top (bottom) panel shows the case for the SNR in a ISM (CSM), for values of the magnetic field of $B=3,10,20\ \umu$G, assuming Bohm diffusion. The dotted lines show the steady-state solution for electrons by \citet{2007ZirakashviliAharonian} for the different magnetic field strengths.
    \label{fig:ismcsmpmax}}
  \end{figure}

In Fig.~\ref{fig:ismcsmpmax} we compare the maximum particle energy for three different magnetic field strengths ($3$, $10$, and $20\ \umu$G) for the case of a CSM (dashed curves) 
and the ISM model (solid curves). Bohm diffusion is assumed in both cases. For clarity, we plot the energy divided by the magnetic field strength to scale away the loss-free $E_{\rm max} \propto B$ behaviour. 
From Eqs.~\ref{eq:Eemax} and \ref{eq:tsync} we find typical synchrotron loss time-scales of $7\times10^3$~yr for $B=10\ \umu$G, and $2\times10^3$~yr for $B=20\ \umu$G for the ISM models. 
For the ISM model with $B=20\ \umu$G we see a significant difference between the maximum proton and electron energy. For the ISM models with lower magnetic field strengths, synchrotron losses are not very important for the energies reached in the simulations. For the CSM models, however,
the maximum cosmic ray energy at a given shock radius is higher due to the larger shock velocity. The higher energies decrease the synchrotron loss time-scales for electrons.
As can be seen in Fig.~\ref{fig:ismcsmpmax} (both for the $B=10\ \umu$G and the $B=20\ \umu$G case) the maximum electron energy is essentially determined by synchrotron losses for a 
shock radius $R > 4$ pc. 

When electron acceleration operates in a regime where the synchrotron losses roughly balance the  acceleration,  
we can compare $E_{\rm max}$ to the exact asymptotic solutions for the cut-off region of the spectrum in the case of a steady, plane parallel shock, 
as derived analytically by \citet{2007ZirakashviliAharonian}. In our notation:
\begin{eqnarray}
E_{\rm max}=\sqrt{\displaystyle \frac{9 \pi e}{\sigma_{\rm T} B}} \: \frac{m_{\rm e} c V_{\rm s}}{(q+2)}.
\end{eqnarray}
At early times, when $E_{\rm max}$ is age-limited, this solution is not valid. 
Later (typically for a shock radius $R > 3-4$ pc) we find this steady-state result slightly underestimates $E_{\rm max}$ for electrons. 
Asymptotically, our value for the electron $E_{\rm max}$ seems to converge to the steady state result.

\begin{figure}
   \centering
     \includegraphics[width=0.5\textwidth]{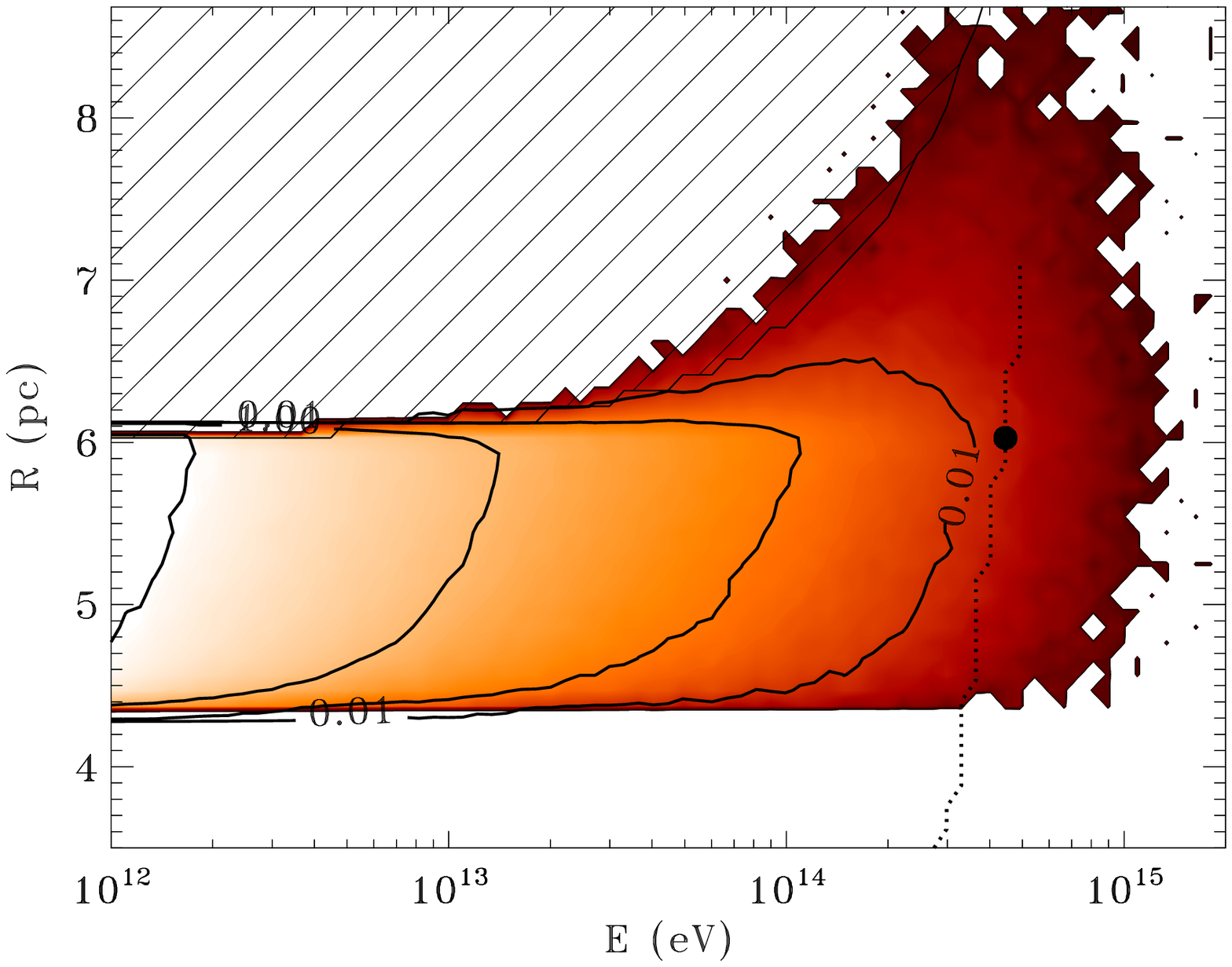}
     \includegraphics[width=0.5\textwidth]{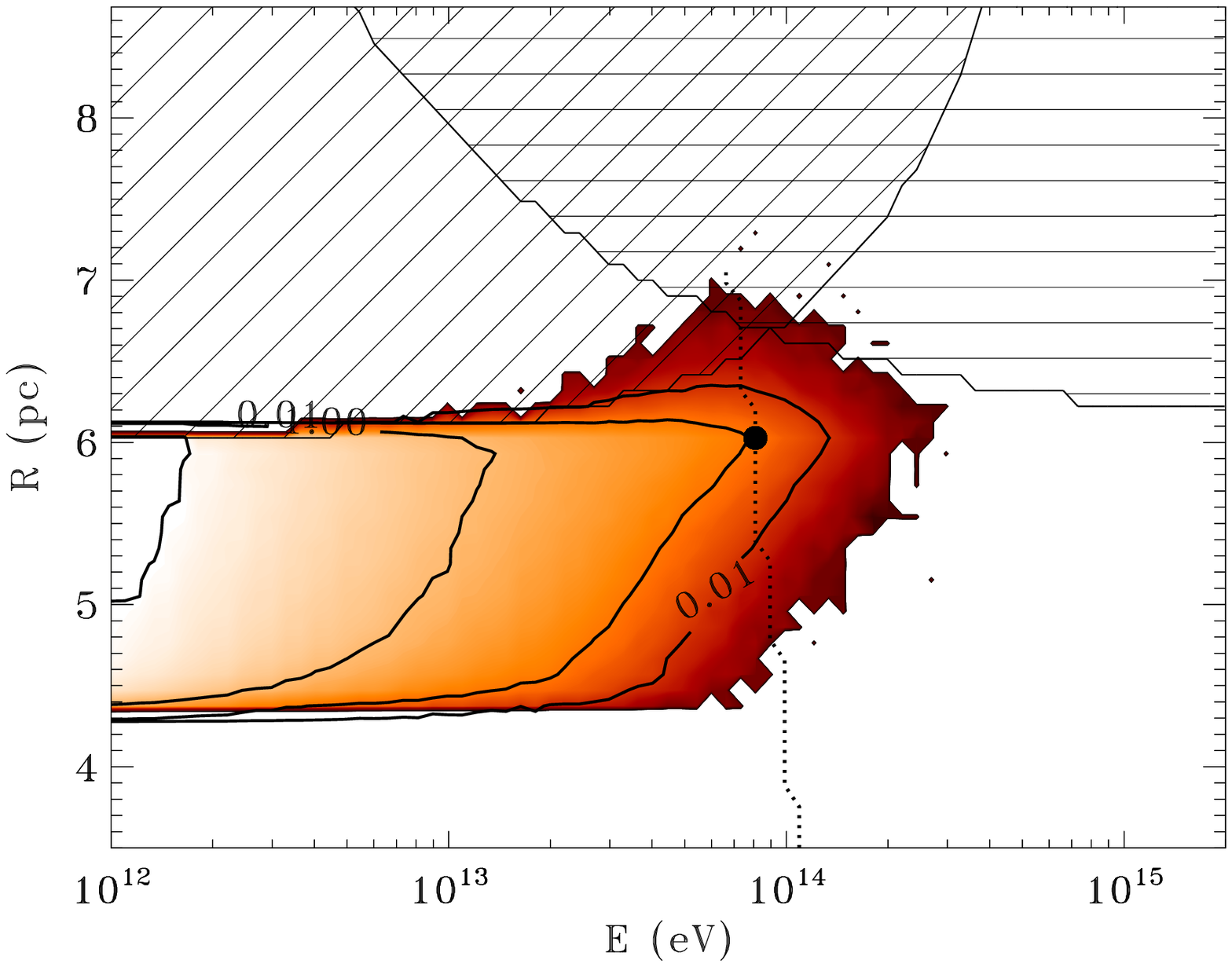}
      \caption[ ] {Cosmic ray protons (top) and electrons (bottom) for the CSM$\kappa_B20$ model for a SNR age of $\sim 1300$~yr. The diagonally shaded area indicates the region $> 4.6$ diffusion lengths upstream of the shock, as calculated with the advection time-scale. The horizontally shaded area shows the region limited by $4.6$ diffusion lengths when the time-scale is determined by synchrotron losses. The dotted line shows the evolution of $p_{\rm max}$ as a function of radius and the thick black dot indicates $p_{\rm max}$ at the shock.
     \label{fig:nxu_norev}}
   \end{figure}

We illustrate the difference between age-limited and loss-limited cosmic ray acceleration in Fig.~\ref{fig:nxu_norev}.
There we show the cosmic-ray distribution of in energy-radius phase-space for a SNR with an age of $\sim 1300$~yr, using the CSM model with $B=20\ \umu$G. 
The relativistic particles are accelerated at the forward shock, where the particles reach the highest energy. 
The maximum energy of protons is limited by the age of the remnant, and therefore still increases at the end of the simulation. 

Sufficiently far downstream from the shock, the `older' cosmic rays are located that have been advected away from the shock. 
The maximum energy of this older population is lower, as reflected in the distribution of particles. 
The dotted line in the figure shows the cut-off energy as a function of radius, 
and the thick black dot marks the intersection of the shock location with the cut-off energy in the energy-radius plane. 
The majority of cosmic rays is located between the forward shock and the 
contact discontinuity. 

Upstream, cosmic rays diffuse ahead of the shock over a typical distance set by the diffusion length $L_{\rm d} \sim V_{\rm s}/\kappa_{\rm B}(E) \propto E$. 
The diagonally shaded area is the region more than $4.6$ diffusion lengths ahead of the shock, in which 99 per cent of the particles should be 
located ($F(p)=F(p)_0\int e^{-\Delta x/x_0}=0.01 F(p)_0$). 
In this model the attainable electron energy is limited mostly by synchrotron losses, and considerably less than the proton energy at an age of $\sim 1300$~yr. 
The high-energy electrons therefore diffuse over smaller distances compared to protons. 
The horizontally shaded region in the electron plot shows the region more than $4.6$ loss-limited diffusion lengths  ahead 
of the shock, ($L_{\rm loss,d} =\sqrt{2 \kappa t_{\rm loss}}$, with $t_{\rm loss}=6 \pi m_{\rm e}^2 c^3/\sigma_T B^2 E$).

\begin{figure}
   \centering
   \includegraphics[width=0.5\textwidth]{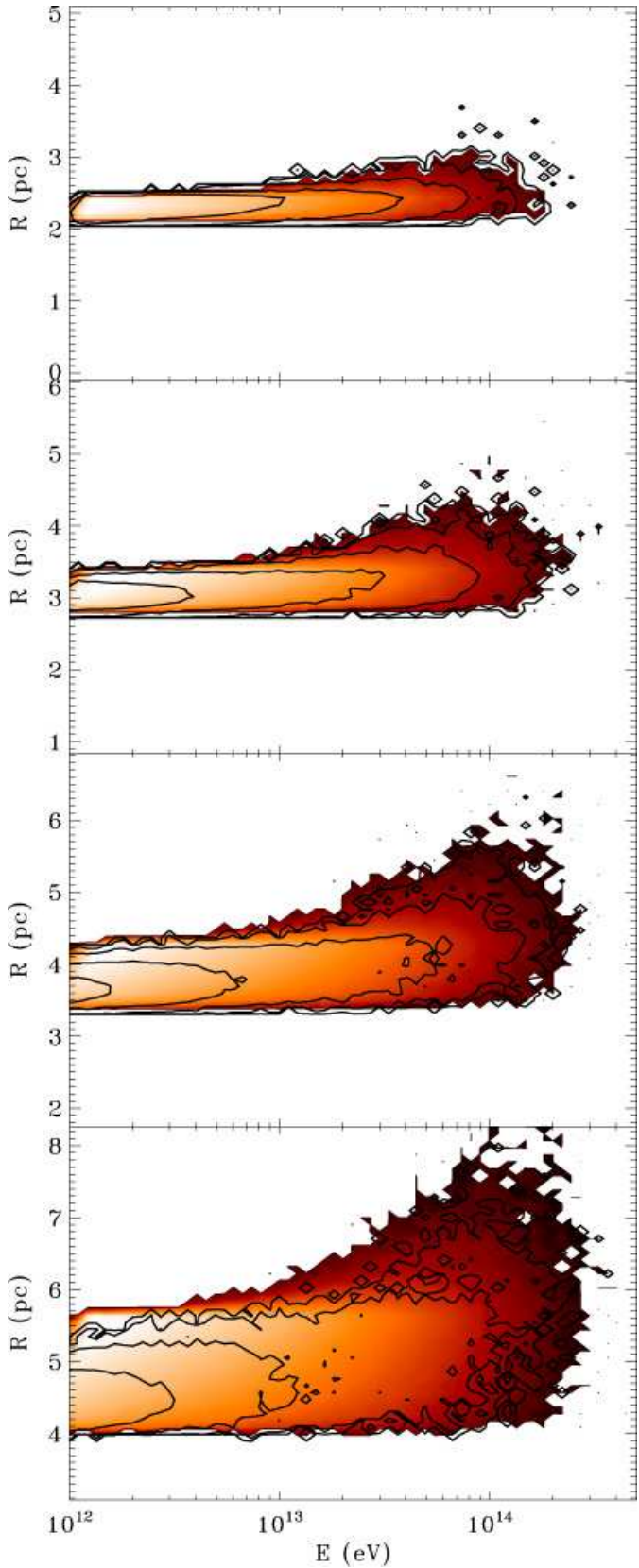}
      \caption[ ] {
       The evolution of the proton distribution in the energy-radius plane as a function of time for the ISM$\kappa_B10$ model. 
The filled contour plots the total particle distribution for times (top to bottom) $381$, $635$, $953$, and $1587$~yr. 
The contour lines track the population of particles that is injected up to time $381$ yr, after which the particles from this early population 
are still subject to diffusion and acceleration. 
      \label{fig:nxu_time}}
   \end{figure}

In Fig.~\ref{fig:nxu_time} we follow the population of protons that were injected during the first 300 yr of the simulation, and track their distribution in the energy-radius plane
as the simulation progresses. 
At the end of the simulation (at $t\approx 1500$~yr), they constitute about 10 per cent of the total proton population. 
The maximum energy in the distribution of these protons turns out to be similar to that of the entire proton poulation. 
This is due to the fact that for this particular model (ISM$\kappa_B$10p), the Sedov-Taylor time is roughly equal to the time at which we start tracking the 
population of particles. As we discussed earlier, the acceleration efficiency is highest around the Sedov-Taylor transition, 
after which the maximum energy of the particles only increases by a factor of a few.

\begin{figure}
 \centering
 \includegraphics[width=0.5\textwidth]{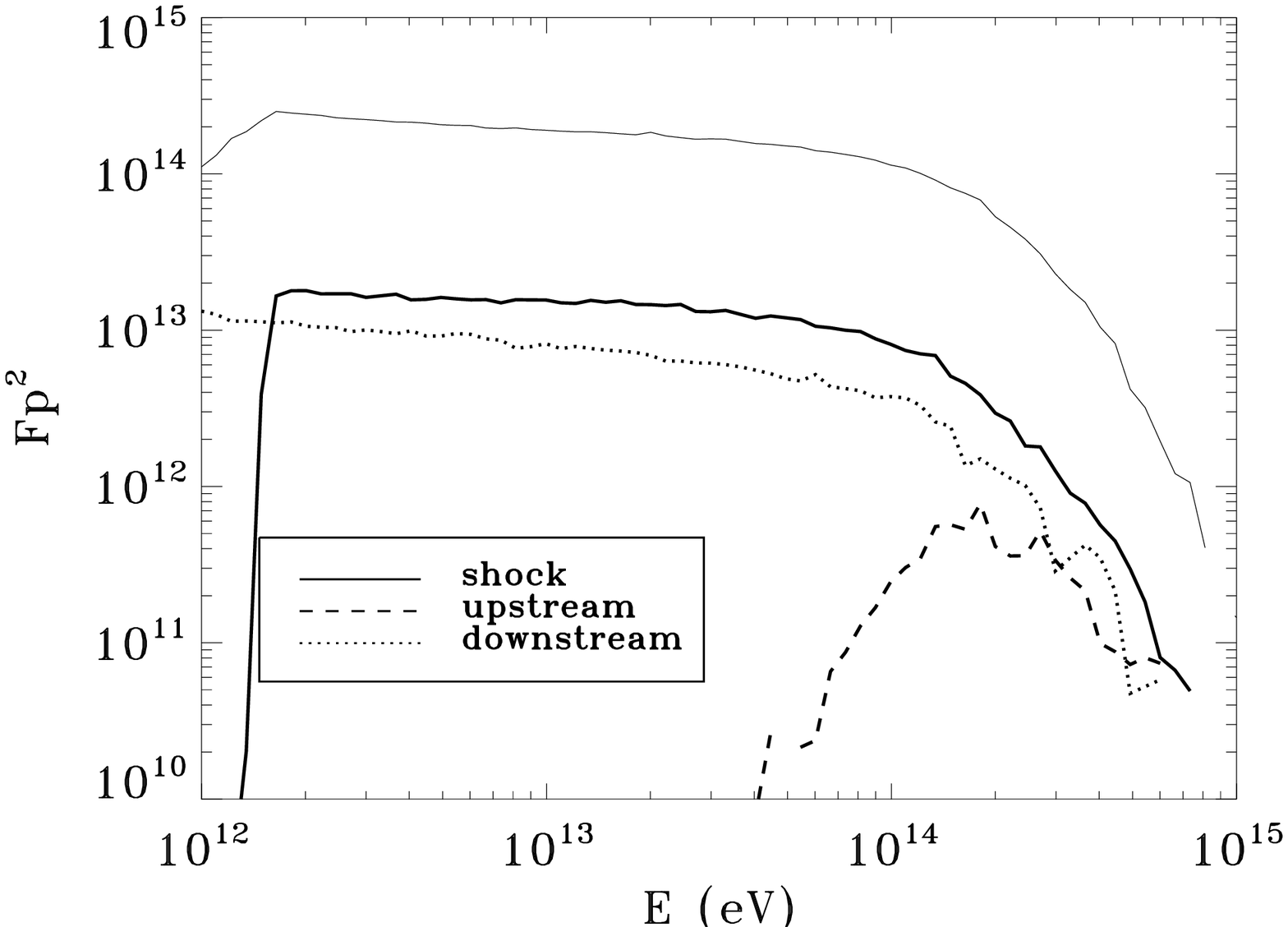}
 \includegraphics[width=0.5\textwidth]{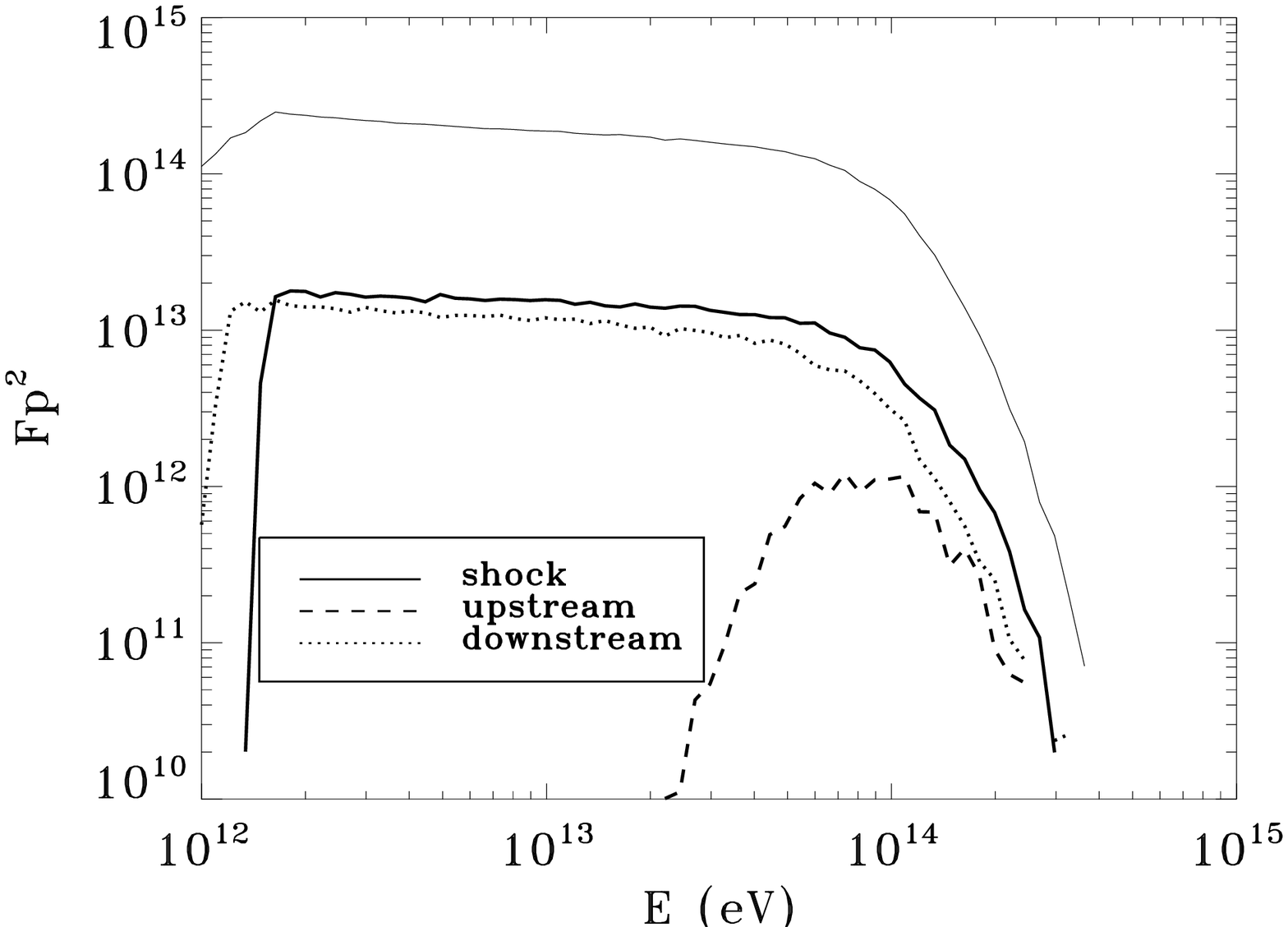}
      \caption[ ] {Proton (top) and electron (bottom) spectra (CSM$\kappa_B10$ model) at the shock and at locations of one diffusion length 
	(for an energy $E_{\rm max}$) upstream and downstream of the shock, with $\Delta x_{\rm diff}=\sqrt{2\kappa t_{\rm adv}}$ and $t_{\rm adv}=\kappa/V_{\rm sh}^2$. 
	The thin solid line shows the cumulative spectrum.
      \label{fig:spectimeloc}}
   \end{figure}

We have determined that the overall spectrum is somewhat steeper than the canonical $q=2$ power law. 
We attribute this to the losses that are not taken into account in the analytical calculations for planar, steady shocks. 
It is therefore interesting to look at the spectrum at the shock, one diffusion length $L_{\rm d}(E_{\rm max})$ upstream and downstream, 
and compare it to the spectrum of all particles in the SNR. 
In Fig.~\ref{fig:spectimeloc} we show these spectra for the CSM$\kappa_B$10 model for electrons and protons. 
It becomes apparent that the spectrum at the shock follows the analytically predicted power law slope quite closely. 
Upstream of the shock only the most energetic particles can be found. Since we probe the region upstream at a diffusion length for $E_{\rm max}$, 
this is where the peak of the upstream spectrum is found. At the same distance downstream, the slope of the spectrum is slightly steeper than at the shock. 
The spectrum is loss-limited for the electrons, as is evident from the sharper cut-off. 

In (semi-)analytical models, the spectrum at the shock in the presence of losses is often described by \citep[e.g.][]{2001MalkovDrury}:
\begin{eqnarray}
F(p) \propto p^{-q} \: {\rm e}^{-\left(p/p_{\rm max}\right)^\alpha},
\end{eqnarray}
with $\alpha=1$ for protons and $\alpha=2$ for electrons. We find that the simulated spectra quite nicely follow this 
exponential cut-off prescription for the cumulative particle spectrum. 
At the shock, the cut-off for protons is slightly sharper than usually assumed: we find $\alpha \approx 1.2-1.3$. 
For electrons, the cut-off is less sharp, but still sharper than for protons: it closely follows $\alpha \approx 1.7$.
This will depend on to what extent the electron spectrum is terminated by synchrotron losses.

\subsection{Re-acceleration at the reverse shock}
\label{sec:rev}

At the forward shock, the magnetic field is likely amplified by the Bell-Lucek mechanism. 
Whether a seed magnetic field with values as low as to be expected at the reverse shock from an expanding SNR would be sufficient to trigger magnetic field amplification, 
and whether this amplification ultimately leads to a field strength that is high enough to confine particles at the reverse shock, 
is still a matter of discussion \citep{2005Ellisonetal}. 
Some studies require higher seed values for the field strength, while other studies observe magnetic field amplification from an almost zero field \citep{2008Changetal}. 
We explore what happens when the magnetic field in the ejecta is strong enough to confine cosmic rays when Bohm diffusion is assumed. For a stronger magnetic field, the particles can undergo many shock crossings because of their smaller mean free path, and effectively can be re-accelerated at the reverse shock. A small mean free path also makes the expansion losses in the flow relatively less important. For simpliciy, we consider the case where the magnetic field strength in the ejecta is equal to that in the CSM/ISM. 

\begin{figure}
   \centering
   \includegraphics[width=0.23\textwidth]{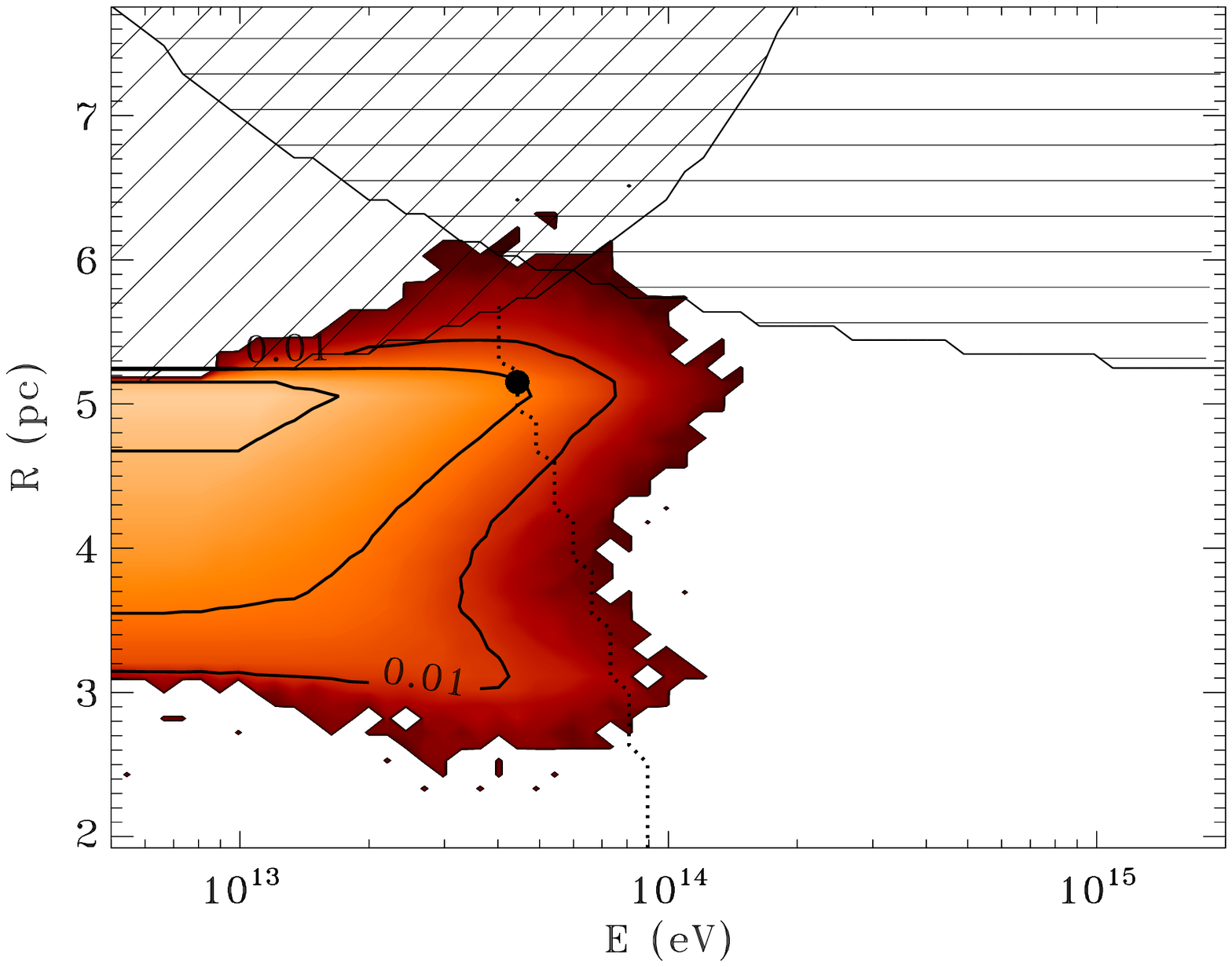} \includegraphics[width=0.23\textwidth]{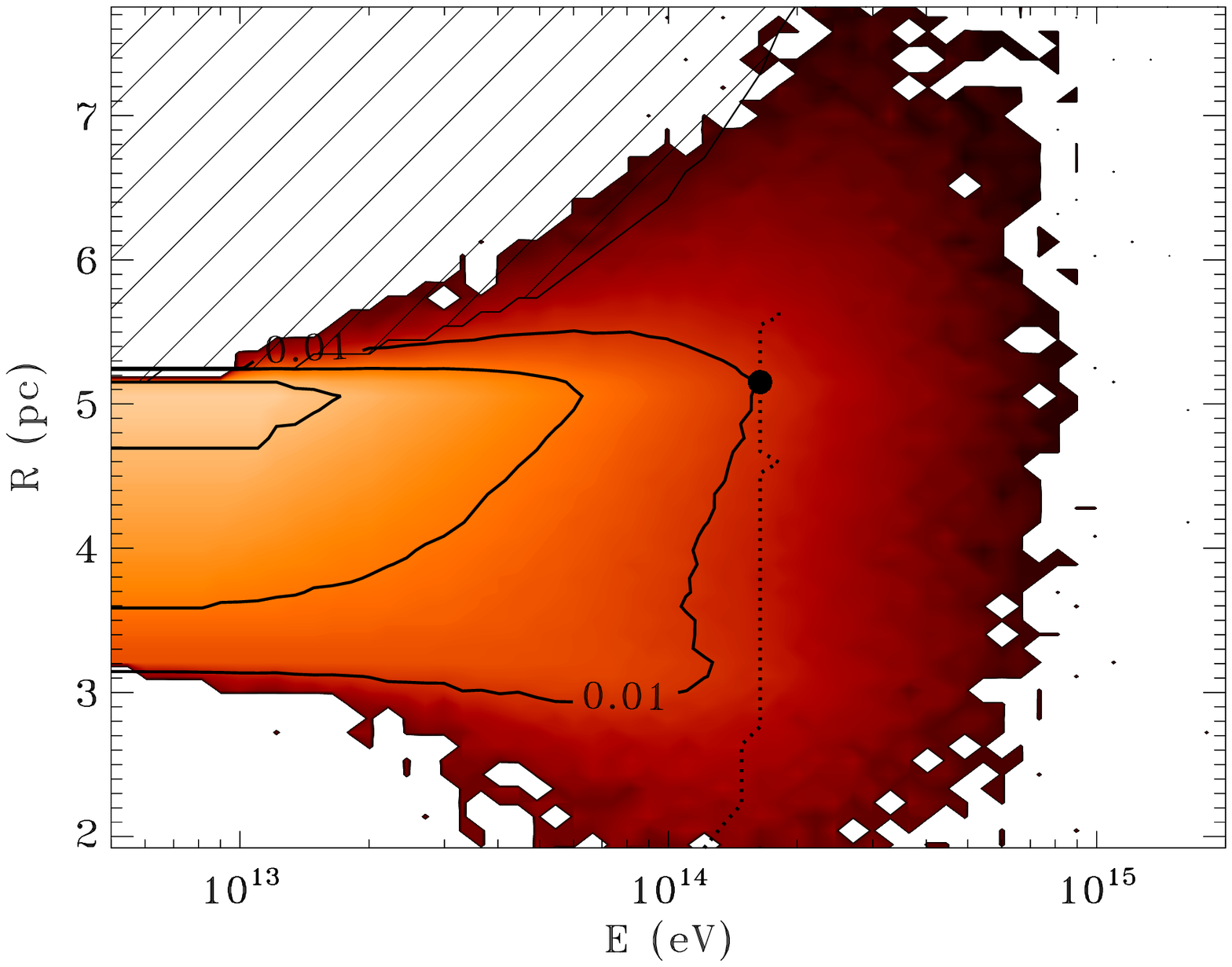}
       \includegraphics[width=0.23\textwidth]{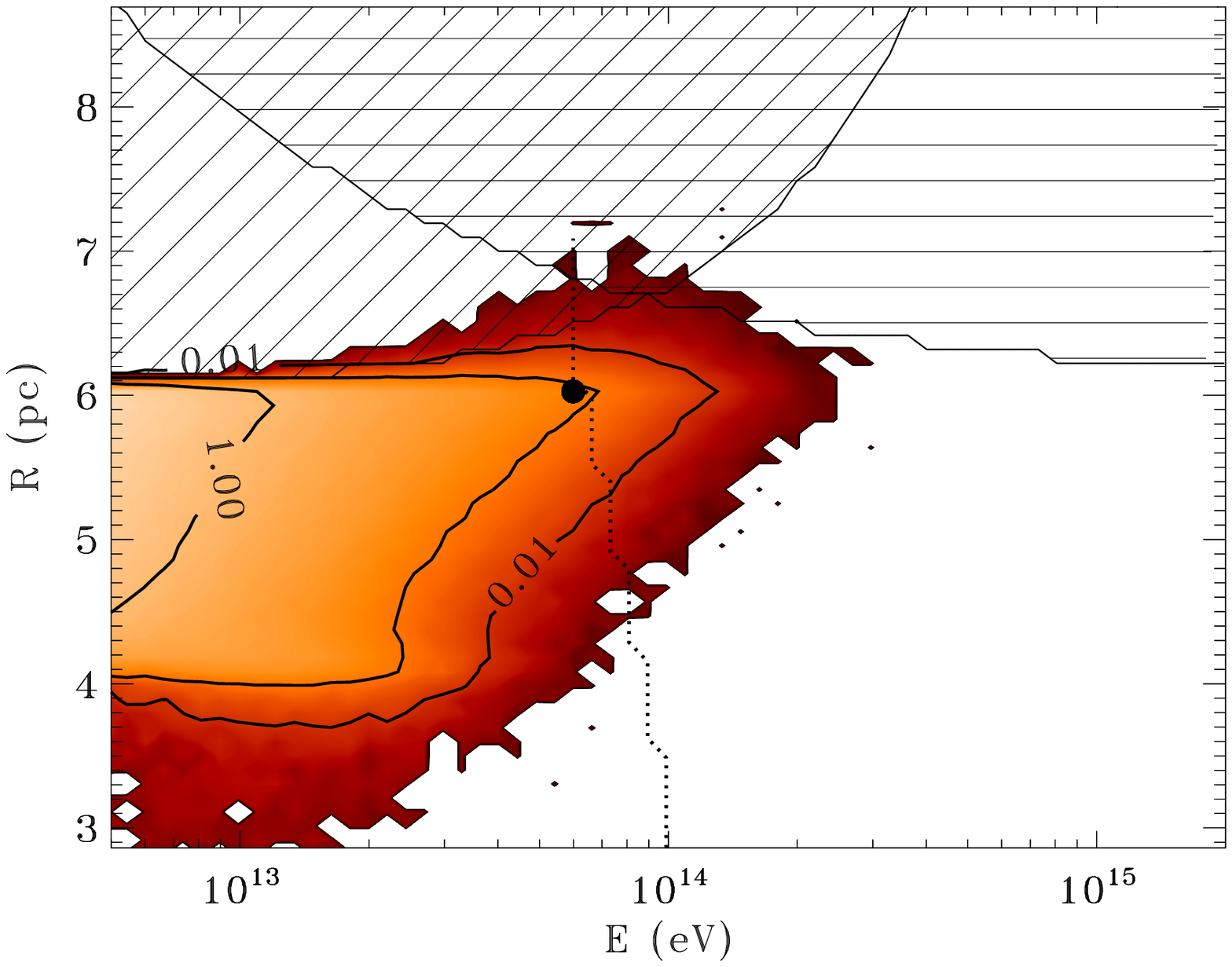} \includegraphics[width=0.23\textwidth]{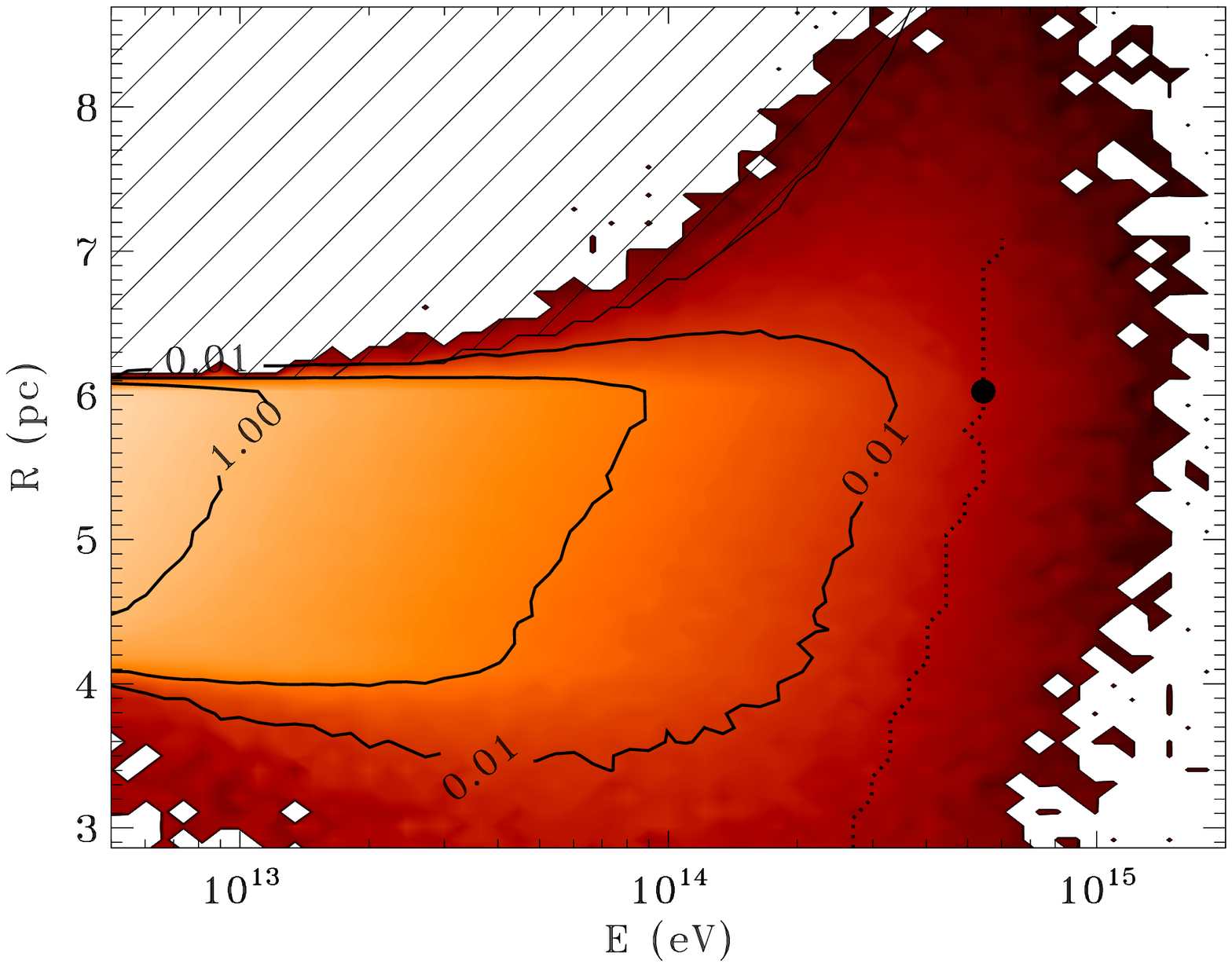}
      \caption[ ] {
The distribution of protons (right) and electrons (left) for a SNR in a ISM (top) or CSM (bottom), when the magnetic field at the reverse shock is amplified 
to the same levels as at the forward shock. We assume a magnetic field strength of $20\ \umu$G, and the distribution is plotted for a remnant of an age of about $700$~yr. 
The location of the shock is indicated at the location of $E_{\rm max}$ with a black dot.
\label{fig:nxu_rev}}
   \end{figure}

In our simulations we observe that very energetic cosmic rays diffuse far enough downstream into the remnant to reach the reverse shock.
Re-acceleration can then occur if the local magnetic field is sufficiently strong.
Fig.~\ref{fig:nxu_rev} shows that the distribution in the energy-radius plane now has two peaks in energy. The highest energy (around $10^{14}$ eV) 
is still attained at the forward shock, 
but acceleration at the reverse shock now causes an additional peak in the energy spectrum at its location ($R \sim 3$ pc for $t \sim 700$~yr).

\begin{figure}
  \centering
  \includegraphics[width=0.5\textwidth]{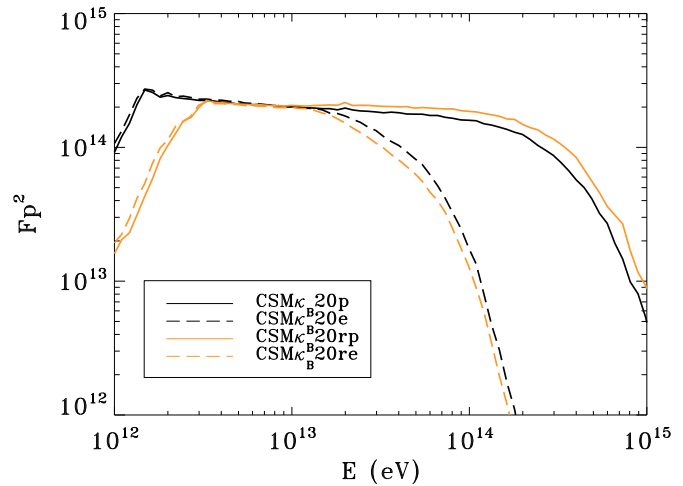}
 \caption[ ] {Spectra when the magnetic field at the reverse shock is equally strong as at the forward shock for the CSM$\kappa_B20$r model (coloured). 
The solid (dashed) lines show the proton (electron) spectrum for a remnant age of $t=1587$~yr. 
The black curves show the spectra when there is no re-acceleration at the reverse shock.
    \label{fig:rev_norevspec}}
  \end{figure}

If magnetic fields turn out to be indeed high enough to confine cosmic rays at the reverse shock, the picture sketched here is over-simplified. 
Apart from the cosmic rays that have diffused far into the SNR from the forward shock, an additional cosmic ray component may arise if higher-Z elements from the supernova ejecta are 
accelerated at the reverse shock. Figure~\ref{fig:rev_norevspec} shows the influence of re-acceleration on the shape of the spectrum.  
The proton spectrum is slightly flatter, and extends to a slightly higher energy. The electron spectrum is hardly modified as synchrotron losses lead to a cut-off around
$10^{13.5}$ eV.

\begin{figure}
  \centering
\includegraphics[width=0.5\textwidth]{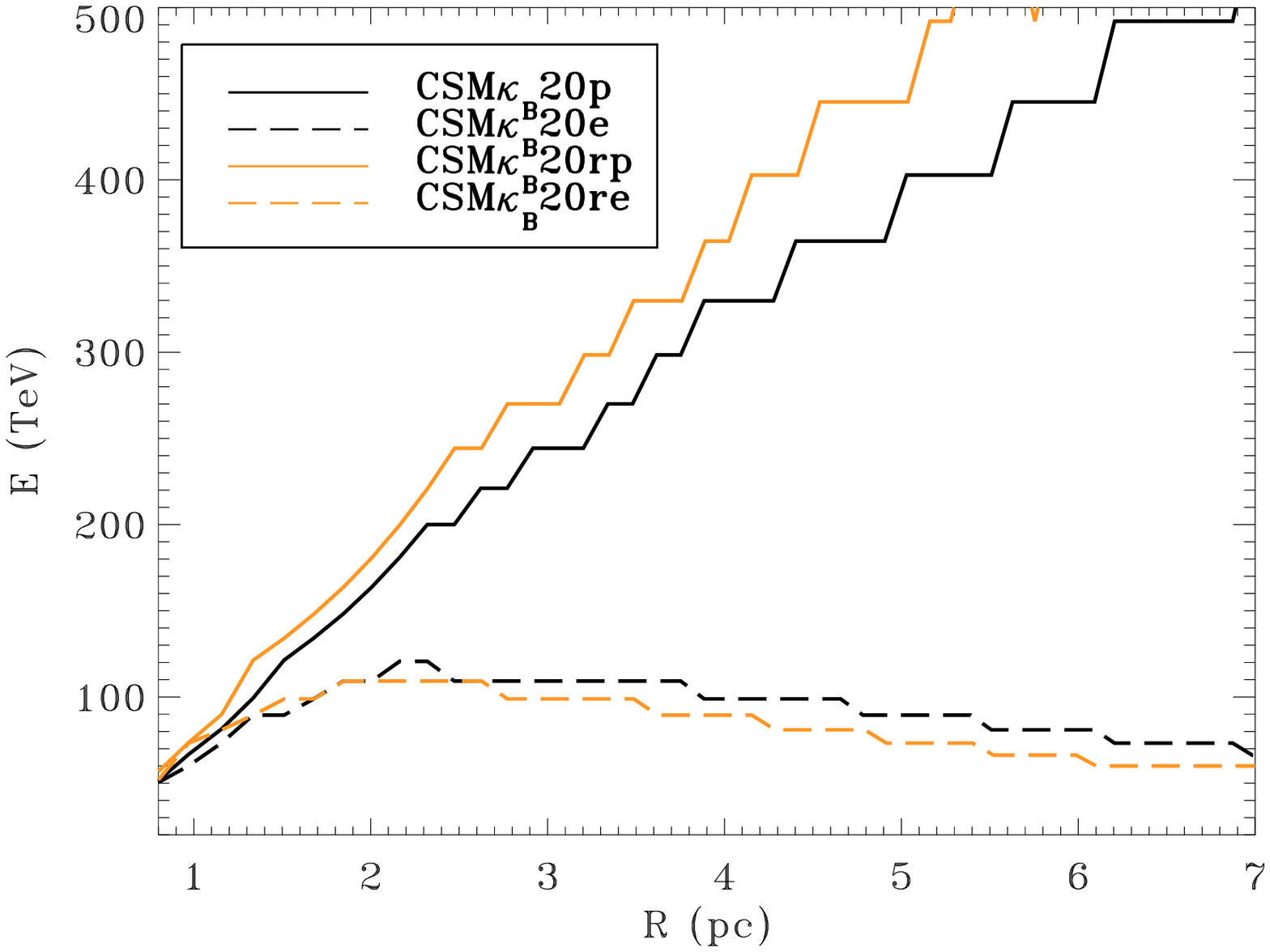}
\includegraphics[width=0.5\textwidth]{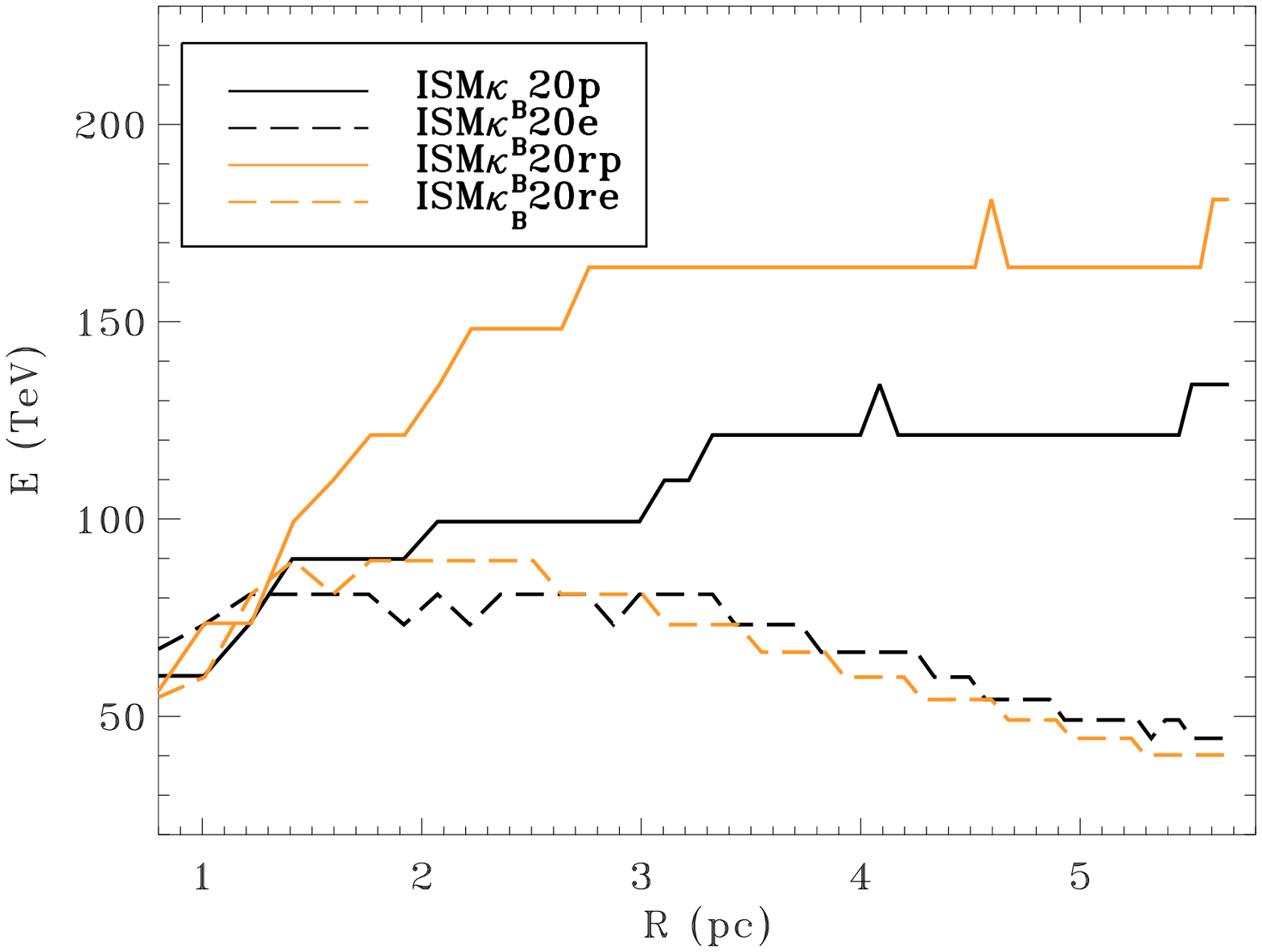}
    \caption[ ] {Maximum energy as a function of time for both relativistic electrons (dashed) and protons (solid) in a SNR where the magnetic field at the reverse shock is $20\ \umu$G, as it is in the rest of the remnant (coloured), compared to the model with negligible magnetic field in the ejecta (black). As in the case of no re-acceleration, the energy the particles obtain in the CSM models (top) is higher than in the ISM models (bottom). While for protons the energy is higher in the `reacceleration' models, for the electrons this turns out mostly not to be the case. Since the added magnetic field induces synchrotron losses also downstream of the contact discontinuity, the maximum energy in these models in fact turns out to be lower when the SNR is in its loss-limited regime.    
   \label{fig:rev_norev}}
  \end{figure}

Figure~\ref{fig:rev_norev} shows that when re-acceleration takes place, the maximum proton energy increases compared to the situation 
without re-acceleration at the reverse shock. For electrons this initially is also the case, as long as the particle energy is limited by the time spent in the source. 
However, in the loss-limited regime, the maximum electron energy is less compared with the case of weak fields near the reverse shock as the larger volume with a 
high magnetic field leads to more synchrotron losses that are apparently not  compensated by  re-acceleration at the reverse shock.  

\section{Discussion and Conclusions}
\label{sec:discussion}

In this paper, we have calculated diffusive shock acceleration through the first-order Fermi mechanism, using stochastic differential equations. 
We treat the cosmic rays as test-particles and follow their acceleration and propagation along with the evolution of a supernova remnant. 
We have extended the model as described by \citet{1992AchterbergKruells} \citep[and used by e.g.~][]{2004vanderSwaluwAchterberg} 
to account for spherical geometry, something that is relevant when using this model to simulate cosmic ray acceleration in supernova remnants. 
The model is set up generically so it can use local magnetic field strengths to calculate the diffusion coefficient of the test-particles. 
However, in this paper we employ a constant magnetic field strength in the ISM/CSM. The ejecta magnetic field is parametrised in such a way that it either decays as expected for an expanding field that is frozen into the plasma, or we fix the magnetic field at the same strength as used for the ISM/CSM. This is because the numerical method that we use is not applicable in 
cases where strong gradients of the diffusion coefficient that may result from gradients in the magnetic field are present. 
The unique set-up of the calculation of the acceleration of the cosmic rays concurrent with the evolution of the SNR allows us to model the time- and location-dependent spectrum more accurately than other models. The disadvantage of our method is that it does not (yet) include feedback of the cosmic ray pressure onto the local plasma.
Our calculations show the following results.

\subsubsection*{Energy spectrum and spectral slope:}

With our method we can accurately model the spectrum of particles, where the slope is close to $q=2$ for an adiabatic index of $5/3$, and to $q=1.5$ for an adiabatic index of $4/3$. 
The slope of the spectrum depends on the location: near the shock the slope of the spectrum is closest to the analytically predicted value for steady and planar shocks, 
while the slope of the cumulative spectrum (all particles in the source) is steeper. The cumulative spectrum is steeper than that at the shock because it also contains particle populations from downstream of the shock, for which the particles were in the acceleration process for a shorter time and hence have lower cut-off energies. 
We also find that in spherical geometry the overall spectrum is slightly steeper when compared to a simulation in slab geometry. The main reason for this is the inclusion of adiabatic expansion, which causes the acceleration process to be slightly less effective: particles lose a fraction of their energy in the downstream part of the shock crossing cycle, reducing the mean energy gain per cycle. The analytical solution for the steady state parallel geometry where adiabatic losses are excluded therefore does not strictly apply in spherical geometry.
The detailed shape of the spectrum also depends on the injection rate of particles as a function of time, which differs in the ISM and the CSM models that we employ.

\subsubsection*{Maximum particle energy:} 

There are additional differences between the results of our approach and those obtained using steady-state analytical models. 
The cut-off energy $E_{\rm max}$ is different from that obtained analytically from the balance between acceleration and losses. 
For protons the $E_{\rm max}$ is determined by the age of the source. The CSM simulations show a higher $E_{\rm max}$ than the analytical estimate, 
whereas for the ISM models the trend is the other way around. 
We attribute this to the difference in cosmic ray age distribution, where, since we assume the injection rate to be proportional to the amount of swept-up mass, 
the fraction of `older' particles, which are accelerated for a longer time, is relatively high in the CSM model and relatively low for the ISM model. 
For Bohm diffusion, the high-energy end of the spectrum shows a distinct cut-off, either caused by the finite age of the SNR or, in the case of electrons,
by synchrotron losses. 

\subsubsection*{Shape of the cut-off in the energy spectrum:}

The shape of the cut-off region for the cumulative spectrum follows quite nicely the quasi-exponential drop that is often assumed. 
However, if one looks at the spectrum in the close vicinity of the shock, the proton spectrum falls off slightly sharper than for the overall proton spectrum, 
while for electrons in the loss-limited regime the cut-off of the spectrum at the shock is more gradual than the cut-off in the overall spectrum. 

\subsubsection*{CSM versus ISM:} 

The strong dependence of the acceleration rate on the shock velocity causes the cosmic ray distribution to become sensitive to the surrounding medium. 
The density profile of the environment into which the supernova explodes determines the shock velocity and its evolution. 
The shock velocity (together with the magnetic field) therefore determines the acceleration rate and the maximum particle energy that can be attained. 
Because in our models the average shock velocity is much higher for a SNR expanding into a CSM, and because the average cosmic ray age in the remnant is larger in the CSM case, $E_{\rm max}$ is larger in our CSM models than in the ISM models. 

This suggests that 
core-collapse SNe in dense environments, such as expected around a red supergiant, may be the most efficient particle accelerators and therefore the dominant 
contributors to cosmic rays up to the ``knee''-energy. In the absence of a significant magnetic field in the ejecta, 
the particles are mostly located between the blast wave and the contact discontinuity. The distance between those is also sensitive to the 
velocity-evolution of the SNR and therefore the surrounding medium, and determines how long electrons are subjected to synchrotron losses. 

\subsubsection*{Re-acceleration at the reverse shock:} 

If the magnetic field is sufficiently amplified at the reverse shock, cosmic rays that are advected away from the blast wave can be re-accelerated at the reverse shock. 
This has important consequences for the maximum energy and the distribution of the cosmic rays. 
The maximum attainable energy for protons becomes significantly higher if re-acceleration at the reverse shock takes place, 
whereas for electrons the additional synchrotron losses in the now strongly magnetised SNR interior can have the opposite effect. 
In reality it is conceivable that due to localized magnetic field amplification, the net effect is a higher maximum energy for electrons, too.

\bigskip
Overall, we conclude that a time-dependent calculation of diffusive shock acceleration in SNRs shows significant differences compared with steady-state plane-parallel analytical models. The environment of the SNR has a large impact on the maximum attainable energy of the cosmic rays. The age distribution of the cosmic rays determines whether a time-dependent approach yields higher or lower maximum attainable energies. 

 \section*{Acknowledgements}
This study has been financially supported by J.V.'s Vidi grant from the Netherlands Organisation for Scientific Research (NWO). This work was sponsored by the Stichting Nationale Computerfaciliteiten (National Computing Facilities Foundation, NCF) for the use of supercomputer facilities, with financial support from the Nederlandse Organisatie voor Wetenschappelijk Onderzoek (Netherlands Organization for Scientific Research, NWO). K.M.S. acknowledges the hospitality of the astronomy department at the University of Florida, where part of this work was performed.

\begin{appendix}
\section{SDEs in spherical geometry}
\label{app:spherical}

In the equation for the distribution function of relativistic particles we need to take account of geometrical effects in order to describe adiabatic losses in a spherical shock 
geometry. 
The advection-diffusion equation (Eq.~\ref{eq:advdiff}) can be written, in spherical coordinates $(R \: , \: \theta \: , \: \phi)$ with axial symmetry
($\partial/\partial \phi = 0$), as:

\begin{eqnarray}
\label{eq:ad2d}
\frac{\partial}{\partial t} F(R \: , \: \theta \: , \: p \: , \: t)  &=& -S_R-S_\theta-S_p,
\end{eqnarray}
with
\begin{eqnarray}
S_R=\frac{1}{R^2}\frac{\partial}{\partial R}\left[R^2\left(V_R F-\kappa \frac{\partial F}{\partial R}\right)\right],
\end{eqnarray}
\begin{eqnarray}
S_\theta=\frac{1}{R \sin \theta}\frac{\partial}{\partial \theta}\left[\sin \theta \left(V_\theta F-\frac{\kappa}{R} \frac{\partial F}{\partial \theta}\right)\right],
\end{eqnarray}
and
\begin{eqnarray}
S_p=\frac{\partial}{\partial p}\left[\frac{d p}{dt} F\right],
\  {\rm
with} \quad
\frac{d p}{dt}= - \frac{p}{3} ({\bf \nabla \cdot V})
\end{eqnarray}
and
\begin{eqnarray}
	{\bf \nabla \cdot V} = \frac{1}{R^2} \frac{\partial}{\partial R} \left( R^2 \: V \right) + \frac{1}{R \: \sin \theta} \frac{\partial}{\partial \theta} \left( \sin \theta \: V \right) \; .
\end{eqnarray}
We assume isotropic spatial diffusion: ${\bf \kappa}=\kappa {\bf I}$, with ${\bf I}$ the unit matrix in configuration space. 
The magnetic field orientation is not taken into account for the diffusion rate, and diffusion in momentum space (Fermi-II acceleration) is neglected. 

In order to be able to apply the It\^o method (Eq.~\ref{eq:ito}), we have to re-order the differential operators in $S_R$ and $S_\theta$, 
such that they conform to the Fokker-Planck standard form (cf. Eq.~\ref{eq:fokkerplanck}).
We substitute $\tilde F=R^2\ F$ into Eq.~\ref{eq:ad2d} to get:
\begin{eqnarray}
\frac{\partial \tilde F}{\partial t} = -\tilde S_R-\tilde S_\theta-\tilde S_{p}.
\end{eqnarray}
The operator of $\tilde S_R=R^2 S_R$ can be rewritten to the standard form as follows:
\begin{eqnarray}
\label{eq:Roperator}
\tilde S_R&&=\frac{\partial}{\partial R} \left[ V_R \tilde F- R^2 \kappa \frac{\partial}{\partial R}   \left( \frac{\tilde F}{R^2}  \right) \right] \nonumber \\
&& \\
&&=\frac{\partial}{\partial R}\left[\left(V_R +\frac{1}{R^2}\frac{\partial (R^2 \kappa )}{\partial R}\right)\tilde F- \frac{\partial}{\partial R}\left(\kappa \tilde F \right)\right].
	\nonumber 
\end{eqnarray}
This now conforms with the standard form with an effective radial velocity
\begin{eqnarray}
V_R^{\rm eff}=V_R+\frac{1}{R^2}\frac{\partial (R^2 \kappa)}{\partial R}.
\end{eqnarray}
This is the velocity that enters the equivalent SDE.

In a similar fashion, $\tilde S_\theta=R^2 S_\theta$ can be rewritten by changing the independent variable 
to $\mu = \cos\theta$ (so that $\partial/\partial \theta=-\sin \theta \partial/\partial \mu$):
\begin{eqnarray}
\tilde S_\theta=&&\frac{1}{R \sin \theta}\frac{\partial}{\partial \theta}\left[\sin \theta \left(V_\theta \tilde{F}-\frac{\kappa}{R} \frac{\partial \tilde{F}}{\partial \theta}\right)\right] \nonumber \\
	&& \nonumber \\
=&&-\frac{\partial}{\partial \mu} 
\left[ \frac{V_\theta \sin \theta}{R} \tilde F+ \frac{\kappa \sin^2 \theta}{R^2}\frac{\partial \tilde F}{\partial \mu}  \right] \nonumber \\
& & \\
=&&-\frac{\partial}{\partial \mu} 
\left[ \left( \frac{V_\theta \sin \theta}{R} +\frac{2 \mu \kappa}{R^2} -\frac{(1-\mu^2)}{R^2}\frac{\partial \kappa}{\partial \mu} \right) \tilde F \right. \nonumber \\
&&+ \left. \frac{\partial}{\partial \mu}\left(\frac{(1-\mu^2)\kappa  \tilde F}{R^2}\right)  \right]. \nonumber
\end{eqnarray}
In short, this gives us:
\begin{eqnarray}
\tilde S_\theta=\frac{\partial}{\partial \mu} 
\left[ V_\mu^{\rm eff} \tilde F - \frac{\partial}{\partial \mu}\left( D_\mu \tilde F\right) \right]
\end{eqnarray}
with
\begin{eqnarray}
V_\mu^{\rm eff} = -\frac{V_\theta \sin \theta}{R}+\frac{\partial}{\partial \mu}D_\mu \nonumber
\end{eqnarray}
and
\begin{eqnarray}
D_\mu = \left(\frac{1-\mu^2}{R^2}\right) \kappa = \frac{\kappa \sin^2 \theta}{R^2}.\nonumber
\end{eqnarray}
For $\tilde S_p$ we substitute $u=\ln (p/mc)$ and $dp=p\ du$ to obtain the equivalent equation for $u$:
\begin{eqnarray}
\tilde S_u=p \tilde{S}_{p} = -\frac{\partial}{\partial u}\left[\left(\frac{1}{3} {\bf \nabla \cdot V} +\lambda_s B^2 \sqrt{1+e^{2 u}}\right) \tilde F\right].
\end{eqnarray}
We have added on the right-hand side the effect of synchrotron losses, with $\lambda_s$ given by Eq.~\ref{eq:betas}. 
As stated before: synchrotron losses may be neglected for protons but are important for electrons.

We now have the advection-diffusion equation in the standard form of the Fokker-Planck equation to which we can 
apply the It\^o method to calculate the particle distribution function in a stochastic manner (Eq.~\ref{eq:ito}).
We solve the following equations numerically to update the position of the particles in phase space:
\begin{eqnarray}
{\rm d}u&=&-\frac{1}{3}({\bf \nabla \cdot V}) \: {\rm d}t - \lambda_s B^2 \sqrt{1+e^{2 u}} \: {\rm d}t \; , \nonumber \\
& & \nonumber \\
{\rm d}R&=&V_R^{\rm eff} \: {\rm d}t +\sqrt{2 \kappa {\rm d}t}\: \xi_R \nonumber \\
 &=&\left( V_R+\frac{1}{R^2}\frac{\partial (R^2 \kappa)}{\partial R}\right) {\rm d}t + \sqrt{2 \kappa {\rm d}t} \: \xi_R \; , \\
& & \nonumber \\
{\rm d}\mu&=&  V_\mu^{\rm eff} \: {\rm d}t +\sqrt{2 D_\mu {\rm d}t} \: \xi_\mu \nonumber \\
 &=&\left( -\frac{V_\theta \sin \theta}{R}+\frac{\partial D_\mu}{\partial \mu}\right) {\rm d}t + \sqrt{2 D_\mu {\rm d}t}\: \xi_\mu. \nonumber 
\end{eqnarray}
If we revert from $\mu$ to $\theta$ the last equation becomes:
\begin{eqnarray}
R  \: {\rm d} \theta&=&\left( V_\theta+\frac{2 \kappa}{R \tan \theta}+\frac{1}{R}\frac{\partial \kappa}{\partial \theta}\right) {\rm d}t -\sqrt{2 \kappa {\rm d}t}\: \xi_\mu.
\end{eqnarray}
The stochastic terms $\xi_i$ should satisfy $\langle \xi_i \rangle=0$ and $\langle \xi_i \xi_j \rangle=\delta_{ij}$, where $i,j$ stand for $R$ and $\mu$.
Note that these equations solve for $\tilde F=R^2 F$ rather than for $F$.
In a slab geometry with flow velocity $V(x \: , \: t)$ along the shock normal the corresponding equations simplify to Eqn (\ref{1DslabSDE}) of the main paper.

\end{appendix}

\bibliography{../adssample}

\label{lastpage}
\end{document}